\documentclass{article}

\PassOptionsToPackage{numbers,sort&compress}{natbib}

\usepackage[final]{neurips_2023}

\AtBeginDocument{%
  \newgeometry{
    textheight=9in,
    textwidth=6.5in,
    top=1in,
    headheight=12pt,
    headsep=25pt,
    footskip=30pt
  }%
}

\usepackage[utf8]{inputenc}
\usepackage[T1]{fontenc}
\usepackage{url}
\usepackage{booktabs}
\usepackage{amsmath, amssymb, amsfonts}
\usepackage{nicefrac}
\usepackage{microtype}
\usepackage{xcolor}
\usepackage{graphicx}
\usepackage{tabularx}
\usepackage{placeins}

\usepackage{tikz}
\usetikzlibrary{fit, positioning, arrows.meta, backgrounds, calc}

\usepackage[utf8]{inputenc}
\usepackage[T1]{fontenc}

\usepackage{hyperref}
\usepackage{url}
\usepackage{booktabs}
\usepackage{microtype}
\usepackage{xcolor}
\usepackage{color}
\usepackage{setspace}
\usepackage{import}

\usepackage{amsmath}
\usepackage{amsthm}
\usepackage{amssymb}
\usepackage{amsfonts}
\usepackage{nicefrac}
\usepackage{commath}
\usepackage{mathtools}

\usepackage[normalem]{ulem}
\usepackage{comment}
\usepackage{environ}

\usepackage{tikz}
\usetikzlibrary{shapes.geometric, arrows}
\usepackage{subfigure}
\usepackage[percent]{overpic}
\usepackage{caption}

\usepackage{diagbox}
\usepackage{multirow}

\usepackage{algorithm,algpseudocode}

\usepackage{bm}

\def\RR{\mathbb{R}}

\newcommand{\bs}{\boldsymbol}

\newcommand{\diff}[1]{\mathrm{d} #1}

\newcommand{\be}{\begin{equation}}
\newcommand{\ee}{\end{equation}}
\newcommand{\bea}{\begin{eqnarray}}
\newcommand{\eea}{\end{eqnarray}}

\newcommand{\bc}{\begin{center}}
\newcommand{\ec}{\end{center}}

\newcommand{\bi}{\begin{itemize}}
\newcommand{\ei}{\end{itemize}}
\newcommand{\bd}{\begin{description}}
\newcommand{\ed}{\end{description}}

\renewcommand{\chi}{\sigma}

\tikzset{
  latent/.style={circle, draw, minimum size=11mm, inner sep=1pt, line width=0.4pt},
  det/.style={circle, draw, double, minimum size=11mm, inner sep=1pt, line width=0.4pt},
  obs/.style={circle, draw, fill=black!12, minimum size=11mm, inner sep=1pt, line width=0.4pt},
  const/.style={rectangle, draw, fill=black!12, minimum size=8mm, inner sep=3pt, line width=0.4pt},
  point/.style={rectangle, draw, minimum size=8mm, inner sep=3pt, line width=0.4pt},
  plate/.style={rectangle, draw, dashed, inner sep=5mm, rounded corners=2pt, line width=0.4pt},
  edge/.style={-{Latex[length=2mm,width=1.4mm]}, line width=0.4pt}
}

\newcommand{\SP}{\mathrm{SP}}
\newcommand{\E}{\mathbb{E}}
\newcommand{\KL}{\mathrm{KL}}
\newcommand{\softmax}{\mathrm{softmax}}

\usepackage{hyperref}

\title{Hierarchical Bayesian physics-augmented neural networks for populations of hyperelastic laws: specimen variability and uncertainty from scarce, noisy data}

\author{Vincent C.~Scholz$^{a}$, Phaedon-Stelios Koutsourelakis$^{a,b,}$\thanks{Corresponding author.} \\
$^a$ Technical University of Munich, Professorship of Data-driven Materials Modeling, \\
School of Engineering and Design, Boltzmannstr.~15, 85748 Garching, Germany \\
$^b$ Munich Data Science Institute (MDSI), Garching, Germany \\
\texttt{\{vincent.scholz, p.s.koutsourelakis\}@tum.de}
}

\begin{document}

\maketitle

\begin{abstract}
The hyperelastic law of a material such as soft tissue differs between specimens, and often only a few specimens can be tested. Physics-augmented neural networks (PANNs) learn such laws from data without a prescribed functional form while keeping them physically admissible, and Bayesian versions of related models quantify the uncertainty of the law of one specimen or of a pooled group of specimens. A pooled fit, however, attributes the differences between specimens to measurement noise or to uncertainty about a mean law, and neither describes a new specimen. We first develop a Bayesian PANN for one specimen, whose priors keep the learned law plausible for few, noisy data. We then extend it to a hierarchical Bayesian PANN, a nonlinear mixed-effects model in which all specimens share learned polyconvex features and each specimen is a draw from a population of laws. Features, specimens and population are inferred jointly by variational inference, with a separate model of the measurement noise, so that the variability between specimens is not mistaken for noise, the population carries its own uncertainty, and every specimen drawn from it is polyconvex. The population predicts the response of new specimens with credible intervals, without data or from a partial test, and assigns them to cohorts with quantified uncertainty, all without a prescribed form of the law. On synthetic populations, it recovers the population from few specimens and predicts new specimens more reliably than three published methods. On Treloar’s rubber and on healthy and aneurysmal aortas, it reaches the accuracy of a deterministic PANN and assigns no aorta specimen to the wrong cohort with confidence, although it underestimates the uncertainty of untested curves.

\end{abstract}


\clearpage
\section{Introduction}
\label{sec:intro}

To choose an adequate treatment for an aortic aneurysm, physicians have to tell diseased from healthy vessel wall, and the mechanics of the tissue, which differ markedly between the two, can help to do so \cite{niestrawska2016microstructure}. The reuse of recycled plastics raises the same kind of question: a component made from a recyclate can only be designed if its deformation under load can be predicted, and a recyclate is a mixture of material streams in varying proportions \cite{scholz2026thermoplastic}. Both questions depend on the same quantity, the constitutive law of the material, which relates how the material is deformed to the stresses it carries. In both examples the law cannot be looked up, since seemingly similar materials, such as two batches of a recyclate or the tissue of two patients, have different laws, which may differ even in the qualitative shape of their stress--stretch curves. We call these differences between specimens of one material \emph{specimen variability}. The standard remedy is to test the material and to infer its law from the measurements, which is known as constitutive model discovery. We first review how the law is inferred, and return to specimen variability afterwards.

\subsection{Inferring material laws from data}

Approaches to constitutive-model discovery can be ordered by how strongly they restrict the form of the law in advance, which in turn sets how much data they need. At one end, an expert selects a closed-form law, such as the Neo-Hookean, Ogden or Holzapfel-type models \cite{holzapfel2000nonlinear}. If the choice is right, few parameters have to be inferred, the law satisfies the physical requirements, and a single test may suffice to determine it. If the choice is wrong, the result suffers from a model-form error, i.e.\ a discrepancy between the chosen law and the true one, and no amount of data repairs it \cite{scholz_illusion_2026}. For the materials above, the right choice is not known. At the other end, a neural network \cite{ghaboussi1991knowledge} or a Gaussian process \cite{fuhg2022local} maps strains directly to stresses with no restriction beyond generic smoothness. It can approximate any response, but it needs data that cover many deformation modes, and nothing guarantees that its predictions of unseen responses are physically meaningful. The approaches discussed next lie between these two ends.

We focus on hyperelastic materials, such as soft tissue under non-destructive, quasi-static loading. Such  constitutive laws are physically admissible if the energy density is objective and respects the material symmetry, the reference state is stress-free, and boundary value problems with it have solutions, for which polyconvexity is the most widely used sufficient condition \cite{ball1977convexity}. The approaches between the two ends of the spectrum differ in how they ensure these conditions. A hand-selected law satisfies them by design. Library-based methods ensure them term by term, by letting the data choose a non-negative combination of admissible candidate terms: EUCLID finds a sparse combination of invariant-based terms from full-field data \cite{flaschel2021unsupervised}, similar to variational system identification \cite{wang2021inference}, constitutive artificial neural networks (CANNs) learn the weights of terms of fixed form, e.g.\ powers or exponentials of invariants \cite{linka2023new}, and Material Fingerprinting matches the measured response against a database of candidate laws \cite{flaschel2026material,flaschel2026unsupervised,flaschel2026adaptive}. The result is a sparse, interpretable formula, but only within the library: a law outside its span, e.g.\ one that couples invariants multiplicatively, cannot be found. Symbolic regression \cite{abdusalamov2023automatic}, agents built on large language models \cite{tacke2025constitutive} and spline interpolants of the energy \cite{wiesheier2024versatile} remove the library, but in general also the guarantee. Physics-augmented neural networks (PANNs) \cite{klein2022polyconvex,linden2023neural} ensure admissibility by their architecture: invariants as inputs give objectivity and material symmetry, an input convex neural network (ICNN) \cite{amos2017input} with non-negative weights gives polyconvexity, and correction terms give a stress-free reference state. Within these constraints, the network learns the terms itself and can combine invariants. A model-form error nevertheless remains possible if the chosen invariants do not describe the material or the true law lies outside the class of the ICNN, e.g.\ because it is not polyconvex or not a convex, non-decreasing function of the invariants. The PANN thus avoids the choice of a law and still guarantees a polyconvex one, which in our view makes it a good balance for the materials above.

The flexibility of the PANN, however, leaves more to the data, and finite, noisy data do not single out one law. Different parameters can produce the same response, already because of symmetries such as the permutation of hidden units, the non-convex loss makes the result depend on the initialisation, which the original CANN and PANN works handle by selecting the best of several restarts \cite{linka2023new,linden2023neural}, and a flexible model can fit the measurement noise instead of the law. A single fit hides this ambiguity. Bayesian approaches quantify it instead: the posterior assigns a probability to every set of parameters compatible with the data, and thereby a distribution over the responses of the material. Bayesian-EUCLID \cite{joshi2022bayesian} does so with a sparsity-promoting prior on the coefficients of its library, the Bayesian CANN \cite{linka2025} with a variational posterior on the weights of its network, and a sparsified input convex network has been combined with Stein variational gradient descent \cite{padmanabha2024improving}. What these works estimate, however, is a mean law of the material with its uncertainty, for one specimen or for a group of specimens.

\subsection{From a mean law to specimen variability}
\label{sec:intro-pooling}

\begin{figure}[t]
\centering
\includegraphics[width=\textwidth]{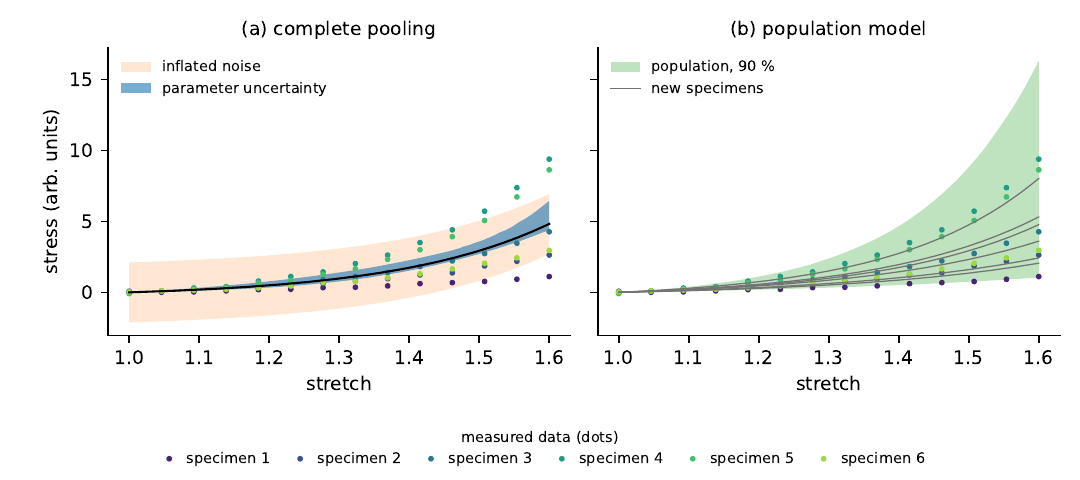}
\caption{The pooling trap (Section~\ref{sec:intro-pooling}), illustrated on a toy problem with six specimens. (a) A pooled fit has no place for the spread of the specimens, so it attributes the spread to the uncertainty of the mean law and to a larger measurement noise. (b) A population model (Section~\ref{sec:hier}) describes the spread by the population from which the specimens are drawn. The two bands in (a) together can look as plausible as the band in (b), but they describe something different. The band of the mean law shrinks as more measurements come in, whereas the population band keeps the spread of the specimens. A draw from the pooled fit is the mean curve with independent noise at every point, whereas a draw from the population is the curve of a specimen.}
\label{fig:pooling}
\end{figure}

A group of specimens is commonly \emph{pooled}, i.e.\ all measurements are treated as repeated measurements of one law. This answers the question of the mean law, but its uncertainty is easily mistaken for specimen variability, which we call the \emph{pooling trap}. A pooled fit attributes the differences between specimens either to parameter uncertainty, which shrinks as measurements are added while the differences do not, or to measurement noise, which perturbs each point independently, whereas a specimen deviates from the mean law coherently along its whole curve and across load cases (Figure~\ref{fig:pooling}). The Bayesian CANN \cite{linka2025} pools the specimens of a tissue group, and its authors note that their differences are absorbed by the noise and the weight uncertainty, consistent with its aim of a mean law per group. Yi et al.\ \cite{yi2025single} and Anton et al.\ \cite{anton2026uncertainty} likewise merge the scatter between samples with the measurement noise into one pointwise error term.

\subsection{Requirements for a population model}
\label{sec:intro-req}

The pooling trap is avoided by a hierarchy over the specimens: each specimen
has its own law, and these laws are draws from a population whose mean and
spread are inferred. For the applications above, such a population model is
chosen, calibrated on the tested specimens and used for new ones. We state the
requirements that we consider the most important for each of these stages. The
\textbf{M}odel has to satisfy:
\begin{enumerate}
  \renewcommand{\theenumi}{M\arabic{enumi}}
  \renewcommand{\labelenumi}{\theenumi}

  \item \label{req:law} \textbf{No law chosen in advance.} The data should
  determine the form of the law, so that materials without an established law
  can be described (Section~\ref{sec:intro}).

  \item \label{req:adm} \textbf{Admissible draws.} Every draw should be a
  polyconvex law with a stress-free reference state, so that drawn laws can
  enter finite element simulations directly. Discarding inadmissible draws
  would change the population that a study over many draws samples.

\end{enumerate}
The \textbf{C}alibration on the tested specimens, which are often scarce, only
partly tested and corrupted by noise, has to provide:
\begin{enumerate}
  \renewcommand{\theenumi}{C\arabic{enumi}}
  \renewcommand{\labelenumi}{\theenumi}

  \item \label{req:data} \textbf{Affordable data.} The population should be
  inferred from a handful of specimens, some of them only partly tested.

  \item \label{req:noise} \textbf{Noise separated from the material spread.} The
  spread between specimens should be separated from the measurement noise
  (Section~\ref{sec:intro-pooling}), so that the population describes the
  material and not the testing device.

  \item \label{req:unc} \textbf{Uncertainty about the population.} The
  uncertainty about the mean law and the spread should be reported, so that a
  population from five specimens is not read as definitive.

\end{enumerate}
The \textbf{U}se of the calibrated population for new specimens requires:
\begin{enumerate}
  \renewcommand{\theenumi}{U\arabic{enumi}}
  \renewcommand{\labelenumi}{\theenumi}

  \item \label{req:pred} \textbf{Prediction from a partial test.} The untested
  load cases of a new specimen should be predicted from its tested ones, so
  that a quick test of a new batch suffices for its design.

  \item \label{req:cohort} \textbf{Assignment to a cohort.} A new, possibly
  partly tested specimen should be assigned to a cohort with quantified
  uncertainty, since a hard assignment can be wrong while looking certain.

\end{enumerate}
None of these requirements is new on its own. In the next section, we compare
related work along them.

\subsection{Related work on populations of laws}

Hierarchical models over specimens, i.e.\ nonlinear mixed-effects models
\cite{lindstrom1990nonlinear,gelman2006data}, are well established for
constitutive laws of fixed form, and have been applied, for example, to tendon
and skin \cite{casey2025exploring,pensalfini2023mechano}. Since each specimen
has its own parameters, measurement noise is modelled separately and the
population carries a posterior, they meet the requirements C and U, but the
form of the law is chosen in advance, against \ref{req:law}.

Two works combine a population with learned laws and are therefore the closest
to the proposed method. The Gaussian CANN \cite{mcculloch2025discovering} scales
the term weights of a CANN for each sample by random factors drawn from a joint
Gaussian, whose mean and covariance are learned by maximum likelihood together
with a sparsity-promoting penalty. Ta\c{c} et al.\ \cite{tac2024generative} fit
each specimen with a polyconvex energy built from neural ODEs, of which only the
last layers are specific to a specimen, and then train a score-based diffusion
model on these specimen-specific parameters. Neither separates the measurement
noise from the spread between specimens (\ref{req:noise}): in the Gaussian CANN,
the scatter of the weights is the only source of stress variance, and in the
diffusion model, the specimens are fitted without a noise model, so that the
noise enters their parameters. Both estimate the population as a point and do
not report its uncertainty (\ref{req:unc}). Beyond this, the Gaussian CANN is
restricted to a library of single-invariant terms, of which a sample can only
rescale those that the mean law retains (\ref{req:law} in part), and a draw is
admissible only if all of its weights are non-negative, which is bounded but not
guaranteed (\ref{req:adm} in part). Its loss compares the pointwise mean and
variance of the stress with the data and does not use which points belong to
the same sample, so that the relation between the load cases of a sample is set
by the model rather than learned (\ref{req:pred} and \ref{req:cohort} in part).
The two-stage construction of Ta\c{c} et al.\ enters each specimen into the
density as an exact point, whether its fit is well determined or rests partly
on an extrapolation of an incomplete test, and the authors note that up to $100$
specimens might be needed (\ref{req:data} in part). With few specimens, the
density can be narrower than the population, and since conditioning a new
specimen can only select among the laws that the density contains, this limits
\ref{req:pred}. In contrast to the parametric population used here, however,
the diffusion model can represent skewed or multimodal populations and
spatially heterogeneous fields, so that the two approaches are complementary.

Further works use the variability between specimens without inferring a
population: conformal bands around a polyconvex model trained on many samples
\cite{bahmani2026conformal}, a prior learned from synthetic stress curves of an
assumed closed-form law \cite{yin2023generative}, and deterministic latent or
measured inputs to (partially) input convex networks
\cite{han2026constitutive,klein2023parametrised}. Outside mechanics, random
effects \cite{wortwein2023neural,tran2018bayesian,slavutsky2026neural} and
Bayesian last layers over shared features \cite{snoek2015scalable,harrison2018meta}
have been placed in neural networks, without a notion of an admissible law. To
the best of the authors' knowledge, none of these works infers a population of
admissible learned laws, separated from the noise and with its uncertainty
(Table~\ref{tab:positioning}).

\begin{table}[htbp]
\centering
\small
\renewcommand{\arraystretch}{1.15}
\resizebox{\textwidth}{!}{%
\begin{tabular}{@{}lccccccc@{}}
\toprule
 & \multicolumn{2}{c}{model class} & \multicolumn{3}{c}{calibration} & \multicolumn{2}{c}{use} \\
\cmidrule(lr){2-3} \cmidrule(lr){4-6} \cmidrule(lr){7-8}
 & M1 no law & M2 admissible & C1 affordable & C2 noise & C3 population & U1 partial & U2 cohort \\
method & in advance & draws & data & separated & uncertainty & test & assignment \\
\midrule
Bayesian-EUCLID \cite{joshi2022bayesian} & (\checkmark) & mostly & -- & -- & -- & -- & -- \\
Sparse Bayesian input convex network \cite{padmanabha2024improving} & \checkmark & \checkmark & -- & -- & -- & -- & -- \\
Bayesian CANN \cite{linka2025} & (\checkmark) & -- & -- & -- & -- & -- & -- \\
Mixed effects, fixed law \cite{casey2025exploring,pensalfini2023mechano} & -- & parameters only & \checkmark & \checkmark & \checkmark & \checkmark & \checkmark \\
Gaussian CANN \cite{mcculloch2025discovering} & (\checkmark) & (\checkmark) & \checkmark & -- & -- & (\checkmark) & (\checkmark) \\
Diffusion over laws \cite{tac2024generative} & \checkmark & \checkmark & (\checkmark) & -- & -- & (\checkmark) & -- \\
\midrule
This work & \checkmark & \checkmark & \checkmark & \checkmark & \checkmark & \checkmark & \checkmark \\
\bottomrule
\end{tabular}}
\caption{Comparison of the cited approaches with the requirements \ref{req:law} to \ref{req:cohort} of a population model, grouped into the model class (M), its calibration on the tested specimens (C) and its use for new specimens (U). A tick means met, a tick in brackets met in part as explained in the text, and a dash not met or not addressed. ``Mostly'' means that the library also contains terms that are not polyconvex, and ``parameters only'' that the constraints hold for the parameters of a fixed law rather than for the resulting energy. The entries follow our reading of the cited works.}
\label{tab:positioning}
\end{table}

\subsection{Contributions and outline}

We propose a hierarchical Bayesian PANN. It builds on the PANN of
\cite{linden2023neural}, an input convex neural network over the invariants of
the deformation, and treats all of its parameters as probabilistic. We regard
the outputs of the last hidden layer as probabilistic features of the strain
energy, which are learned from the data (\ref{req:law}). All specimens share
these features, so that the data of all specimens inform them and a partly
tested specimen borrows them from the others (\ref{req:data}). The hierarchy
enters only in how the features are combined: a specimen weights them by a
positive amplitude, which scales its stresses, and by positive shares that sum
to one, which set how much each feature contributes to the shape of its law, so
that every draw is admissible by construction (\ref{req:adm}). The amplitudes
and shares of the specimens are draws from a population in these few
coordinates, whose mean and covariance carry a posterior (\ref{req:data},
\ref{req:unc}). The features, the coordinates of each specimen, the population
and a separate measurement noise are inferred together, in one posterior
(\ref{req:noise}). A new specimen receives its own coordinates, which can be
conditioned on a partial test (\ref{req:pred}) or compared across the
populations of several cohorts (\ref{req:cohort}).

In contrast to the Gaussian CANN, the specimens thus weight learned features
rather than pre-selected terms, and in contrast to the diffusion model,
specimens and population are inferred in one posterior rather than in two
stages. Since inputs and features are normalised, the priors in principle need no
tuning for a new material, and like the PANN it builds on, the model applies to
compressible materials, whereas the population models above are formulated for
incompressible ones.

In summary, the main contributions of this work are:
\begin{enumerate}

  \item A Bayesian PANN for a single specimen. Following
  \cite{padmanabha2024improving}, we formulate the PANN of
  \cite{linden2023neural} in a Bayesian way, with priors on the network that
  keep the learned law plausible when the data are noisy and few.

  \item The hierarchical Bayesian PANN described above, with the
  parameterisation of each specimen by one amplitude and the shares of the
  shared features. To the best of the authors' knowledge, it is the first
  hierarchical formulation of a PANN, and the first population model of learned
  constitutive laws that meets the requirements M, C and U together
  (Table~\ref{tab:positioning}).

  \item Procedures that use the fitted population for a new specimen: a
  prediction without data, its conditioning on a partial test
  (\ref{req:pred}) and its assignment to a cohort (\ref{req:cohort}).

  \item Numerical illustrations on synthetic populations with a known ground
  truth, on Treloar's rubber \cite{treloar1944stress} and on healthy and
  aneurysmal aortas \cite{niestrawska2016microstructure,linka2025}. They compare
  the Bayesian PANN with the deterministic PANN \cite{linden2023neural}, and the
  hierarchical model with the Gaussian CANN \cite{mcculloch2025discovering}, the
  Bayesian CANN \cite{linka2025} and the diffusion model
  \cite{tac2024generative}, each run with its original code.

\end{enumerate}

The remainder of this paper is organised as follows. Section~\ref{sec:prelim} recalls the Bayesian inference and PANN building blocks. Section~\ref{sec:method} presents the hierarchical model, its inference and the predictive procedures. Section~\ref{sec:results} reports the numerical illustrations, Section~\ref{sec:discussion} discusses limitations and extensions, and the appendices collect the prior choices and supplementary studies.

\clearpage
\section{Preliminaries}
\label{sec:prelim}

The method of Section~\ref{sec:method} combines Bayesian inference with the physics-augmented neural network (PANN) of Linden et al.\ \cite{linden2023neural}, which we modify in two places.

\subsection{The physics-augmented neural network}
\label{sec:pann}

In the PANN, an input-convex neural network maps invariants of the deformation to the strain energy, and the stress follows from the energy by differentiation.

\paragraph{Kinematics and stress} With the deformation gradient $\bs{F}$, the right Cauchy--Green tensor $\bs{C} = \bs{F}^\top \bs{F}$ and $J = \det \bs{F}$ \cite{holzapfel2000nonlinear}, a hyperelastic material with strain energy density $\Psi(\bs{C})$ has the second Piola-Kirchhoff stress
\begin{equation}
  \bs{S} = 2\, \frac{\partial \Psi}{\partial \bs{C}}.
  \label{eq:stress}
\end{equation}

\paragraph{Invariants as inputs} Expressing the energy through invariants of $\bs{C}$ makes it objective and respects the symmetry of the material. For isotropic, compressible materials the PANN uses
\begin{equation}
  I_1 = \mathrm{tr}\,\bs{C}, \qquad
  I_2 = \mathrm{tr}(\mathrm{cof}\,\bs{C}), \qquad
  I_3 = \det \bs{C} = J^2, \qquad
  I_1^* = -2J,
  \label{eq:invariants}
\end{equation}
where the additional input $I_1^*$ is needed so that the network can represent negative stresses \cite{linden2023neural,klein2022polyconvex}. Anisotropic materials add further invariants, e.g.\ $I_4$ and $I_5$ of the fibre direction for transverse isotropy, and for incompressible materials $I_3$ and $I_1^*$ are constant and dropped \cite{linden2023neural,holthusen2026benchmarking}. We collect the $n_0$ invariants that enter the network in the vector $\bs{\mathcal{I}}$.

\paragraph{Network} An input-convex neural network \cite{amos2017input} with $L$ hidden layers maps the invariants to the energy,
\begin{subequations}
\label{eq:icnn}
\begin{align}
  \bs{g}^{(1)} &= \SP\big(\bs{W}^{(1)}_{+}\,\bar{\bs{\mathcal{I}}} + \bs{b}^{(1)}\big), \label{eq:icnn-first} \\
  \bs{g}^{(h)} &= \SP\big(\bs{W}^{(h)}_{+}\,\bs{g}^{(h-1)} + \bs{b}^{(h)}\big), \qquad h = 2, \dots, L, \label{eq:icnn-hidden} \\
  \Psi_{\mathrm{NN}} &= \bs{w}_{+}\,\bs{g}^{(L)} = \sum_{j=1}^{N} w_{+,j}\, g^{(L)}_j, \qquad
  \bs{W}^{(h)}_{+} = \SP\big(\bs{W}^{(h)}\big), \quad \bs{w}_{+} = f_+(\bs{w}), \label{eq:icnn-out}
\end{align}
\end{subequations}
where $\SP(u) = \log(1 + \exp(u))$ is the softplus function, applied elementwise, $\bar{\bs{\mathcal{I}}}$ are the standardised invariants \eqref{eq:standardise} below, and $\bs{g}^{(h)} \in \RR^{n_h}$ are the activations of the $n_h$ neurons of hidden layer $h$, with $N = n_L$. The weight parameters $\bs{W}^{(h)} \in \RR^{n_h \times n_{h-1}}$, the biases $\bs{b}^{(h)} \in \RR^{n_h}$ and the weight parameters $\bs{w}$ of the output layer are unconstrained real numbers, and the network uses the non-negative weights $\bs{W}^{(h)}_{+}$ and $\bs{w}_{+}$. The standard PANN takes $f_+ = \SP$ \cite{linden2023neural}; we replace it by a different map in Section~\ref{sec:outlayer}. With non-negative weights and a convex, non-decreasing activation, \eqref{eq:icnn} is an input-convex neural network on polyconvex invariants, which makes $\Psi_{\mathrm{NN}}$ polyconvex, i.e.\ convex in $(\bs{F}, \mathrm{cof}\,\bs{F}, \det \bs{F})$ \cite{linden2023neural,klein2022polyconvex}.

Unlike the standard PANN \cite{linden2023neural}, we standardise the invariants, whose values differ by orders of magnitude,
\begin{equation}
  \bar{\mathcal{I}}_m = \frac{\mathcal{I}_m - \mathcal{I}^0_m}{\Delta\mathcal{I}_m}, \qquad m = 1, \dots, n_0,
  \label{eq:standardise}
\end{equation}
where $\mathcal{I}^0_m$ and $\Delta\mathcal{I}_m > 0$ are the mean and the standard deviation of the $m$th invariant over the training states (alternatives in \ref{app:pann}). The map \eqref{eq:standardise} is affine with positive slope and therefore preserves polyconvexity. It puts all inputs on a common scale of order one, on which the prior of Section~\ref{sec:onespecimen} relies.

\paragraph{Correction terms} Three correction terms $\Psi_{\mathrm{corr}}$ of \cite{linden2023neural}, given in \eqref{eq:corrections}, are added to the network energy, $\Psi = \Psi_{\mathrm{NN}} + \Psi_{\mathrm{corr}}$, so that the energy and the stress vanish in the undeformed state and the energy grows without bound for $J \to 0$ and $J \to \infty$. They preserve polyconvexity. The energy is not guaranteed to be non-negative, and we screen this condition numerically (\ref{app:evaluation}). The stress of the PANN follows from $\Psi$ through \eqref{eq:stress}.

\paragraph{Parameters and features} We collect the weight parameters and biases of the hidden layers in $\bs{\theta} = \{\bs{W}^{(h)}, \bs{b}^{(h)}\}_{h=1}^{L}$ and write the stress of the PANN as $\bs{S}(\bs{C}; \bs{\theta}, \bs{w})$. By \eqref{eq:icnn-out}, the network energy is a non-negative combination of the $N$ activations $g^{(L)}_j$ of the last hidden layer, each of which is itself a polyconvex function of the deformation. We call them the \emph{features} of the network: $\bs{\theta}$ determines which features are available, and $\bs{w}$ how strongly each of them contributes.

\subsection{Bayesian inference}
\label{sec:bayes}

We treat the parameters $(\bs{\theta}, \bs{w})$ of the PANN as random variables. What is known about them before a test, such as a plausible order of magnitude of the stress, is expressed by a prior $p(\bs{\theta}, \bs{w})$. A test yields data $\mathcal{D}$, i.e.\ stresses measured in known deformation states, and the likelihood $p(\mathcal{D} \mid \bs{\theta}, \bs{w})$ is the probability of these data for given parameters, which follows from a model of the measurement noise (Section~\ref{sec:onespecimen}). Bayes' rule combines both into the posterior
\begin{equation}
  p(\bs{\theta}, \bs{w} \mid \mathcal{D}) = \frac{p(\mathcal{D} \mid \bs{\theta}, \bs{w})\, p(\bs{\theta}, \bs{w})}{p(\mathcal{D})} \propto p(\mathcal{D} \mid \bs{\theta}, \bs{w})\, p(\bs{\theta}, \bs{w}),
  \label{eq:bayes}
\end{equation}
where the evidence $p(\mathcal{D})$ normalises the posterior and does not depend on the parameters; we use it in Section~\ref{sec:cohort} to compare cohorts. The posterior
concentrates on the parameters that balance reproducing the measurements against plausibility under the prior.
Since every parameter value defines a law $\bs{S}(\bs{C}; \bs{\theta}, \bs{w})$, the posterior is a distribution over constitutive laws, and its spread is the uncertainty about the law that the finite and noisy data leave.

\clearpage
\section{Methodology}
\label{sec:method}

\begin{figure}[htbp]
\centering
\resizebox{\textwidth}{!}{%
\begin{tikzpicture}[
  box/.style={rectangle, draw, rounded corners=2pt, minimum height=9mm, inner sep=4pt, line width=0.4pt, align=center},
  arr/.style={-{Latex[length=2mm,width=1.4mm]}, line width=0.4pt}]
  \node[box] (inv) at (0,0) {invariants\\$\bar{\bs{\mathcal{I}}}(\bs{C})$};
  \node[box, fill=black!6] (feat) at (4.2,0) {shared features\\$\tilde g_1,\dots,\tilde g_N$\\(hidden layers $\bs{\theta}$)};
  \node[box] (comb) at (8.6,0) {specimen weighting\\$\exp(a_k)\,\bs{\pi}(\bs{z}_k)$};
  \node[box] (out) at (12.4,0) {stress\\$\bs{S}(\bs{C}; \bs{\theta}, \bs{w}_k)$};
  \draw[arr] (inv) -- (feat);
  \draw[arr] (feat) -- (comb);
  \draw[arr] (comb) -- (out);
  \node[box, fill=black!6] (pop) at (8.6,-2.6) {population\\$\bs{w}_k \sim \mathcal{N}(\bs{\mu}, \bs{\Sigma})$};
  \draw[arr] (pop) -- node[right, font=\small] {$\bs{w}_k = (a_k, \bs{z}_k)$} (comb);
  \node[font=\small, text width=3.6cm, align=center] (shl) at (4.2,-2.6) {shared by all specimens,\\with posterior uncertainty};
  \draw[arr, dashed] (shl.north) -- (feat);
  \node[font=\small, text width=3.4cm, align=center] (dfl) at (12.4,-2.6) {different for each\\specimen $k$};
  \draw[arr, dashed] (dfl.north) -- (out);
\end{tikzpicture}}
\caption{Structure of the hierarchical Bayesian PANN, which separates what all specimens share from what differs between them. The hidden layers produce $N$ polyconvex features, shared by all specimens. Specimen $k$ weights the normalised features \eqref{eq:featnorm} by an amplitude $\exp(a_k)$, which sets its stress level, and by shares $\bs{\pi}(\bs{z}_k)$ that sum to one, which set the shape of its response. The specimen parameters $\bs{w}_k = (a_k, \bs{z}_k)$ are draws from a Gaussian population \eqref{eq:population}. Since every draw is a positive combination of polyconvex features, every specimen of the population is polyconvex.}
\label{fig:scheme}
\end{figure}
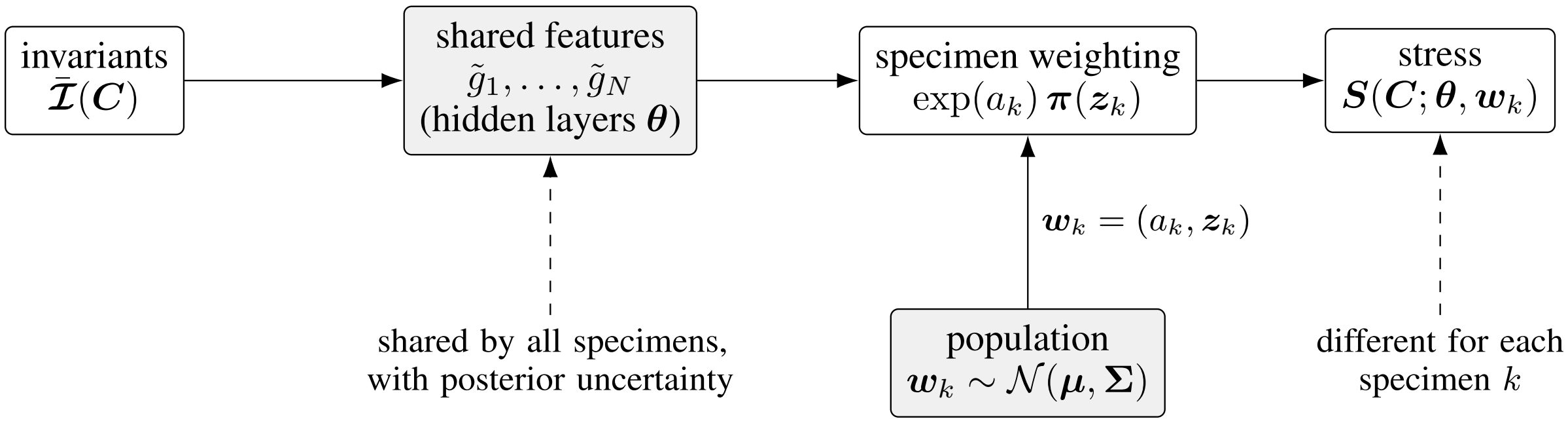

We build a constitutive model for a population of specimens of one material: its mean law, the variability between its specimens, and the uncertainty that remains because only a few noisy tests are available. Section~\ref{sec:onespecimen} turns the PANN of Section~\ref{sec:pann} into a Bayesian model for one specimen. Section~\ref{sec:hier} extends it to a population, in which all specimens share the features of the hidden layers and differ only in how they weight them (Figure~\ref{fig:scheme}). Section~\ref{sec:inference} describes the variational approximation of the posterior, and Section~\ref{sec:prediction} the predictions for new specimens.

\subsection{A Bayesian PANN for one specimen}
\label{sec:onespecimen}

For one specimen, the data are $M$ measurements $\mathcal{D} = \{(\bs{C}_i, \hat{\bs{S}}_i)\}_{i=1}^{M}$ of the stress $\hat{\bs{S}}_i$ in the deformation state $\bs{C}_i$. A test measures only some components of the stress tensor, and the others follow from its conditions, such as stress-free lateral surfaces. We assume that the measured components deviate from the stress of the PANN by independent Gaussian noise, which gives the likelihood
\begin{equation}
  p(\mathcal{D} \mid \bs{\theta}, \bs{w}) = \prod_{i=1}^{M} \mathcal{N}\big(\hat{\bs{S}}_i \mid \bs{S}(\bs{C}_i; \bs{\theta}, \bs{w}), \sigma_\epsilon^2\big),
  \label{eq:lik}
\end{equation}
where we write measurements and stresses as full tensors, and $\mathcal{N}(\hat{\bs{S}} \mid \bs{S}, \sigma_\epsilon^2)$ denotes the product of independent Gaussian densities over the measured components. The noise level $\sigma_\epsilon$, the standard deviation of the measured stress in Pa, is either known, as for a calibrated testing device, or inferred under the prior $\log \sigma_\epsilon \sim \mathcal{N}(\log \sigma_{\epsilon,0}, 1)$. The value $\sigma_{\epsilon,0}$ is a rough estimate of the device precision, and the prior places $\sigma_\epsilon$ within a factor of about $7$ of it with $95\,\%$ probability. The prior on the network parameters should satisfy three conditions: (i) every parameter value gives a polyconvex law, so that every posterior draw does; (ii) the prior is stated before the test and does not depend on the measured stresses, whether these are of order kPa, as for soft tissue, or MPa, as for rubber; (iii) the parameters of the output layer are interpretable, since Section~\ref{sec:hier} places the variability between specimens on them.

\subsubsection{Hidden layers}

The hidden layers $\bs{\theta}$ determine which features are available to the law. We place the fan-in prior \cite{lecun1998efficient}
\begin{equation}
  W^{(h)}_{mn} \sim \mathcal{N}\big(0, \kappa_0^2/n_{h-1}\big), \qquad
  b^{(h)}_{m} \sim \mathcal{N}\big(0, (3\kappa_0)^2\big), \qquad h = 1, \dots, L,
  \label{eq:single}
\end{equation}
on the weight parameters and biases, where $n_{h-1}$ is the number of inputs of layer $h$ and $\kappa_0 = 3$. Since the weights $\SP(W^{(h)}_{mn})$ are non-negative for every draw, condition (i) holds. The two priors have a direct reading for the features. A weight sets how strongly a feature depends on an invariant, or on a feature of the previous layer; with $\kappa_0 = 3$ it ranges over about three orders of magnitude under the prior, from a nearly absent to a dominant dependence. A bias sets the deformation at which a softplus unit turns from flat to linear, i.e.\ where a feature starts to stiffen. Because the invariants are standardised \eqref{eq:standardise}, the tested deformations lie within a few units of zero, and the wide bias prior lets this onset lie anywhere within them, or outside, which leaves the unit linear or switched off over the tested range. The prior thus leaves room for different stiffening behaviours. Its stress level is irrelevant, since the output layer removes it (Section~\ref{sec:outlayer}). Priors following the initialisation of input-convex networks \cite{hoedt2023principled}, and priors with weights summing to a constant, gave no advantage in our experiments.

\subsubsection{Output layer: magnitude and shape}
\label{sec:outlayer}

Stresses range from $\mathcal{O}(10^{3})$\,Pa for soft tissue to $\mathcal{O}(10^{6})$\,Pa for rubber. To satisfy (ii) and (iii), we separate the stress level of a law from the shape of its response,
\begin{equation}
  \Psi_{\mathrm{NN}} = \exp(a) \sum_{j=1}^{N} \pi_j\, \tilde g_j, \qquad \sum_{j=1}^{N} \pi_j = 1, \quad \pi_j > 0.
  \label{eq:outlayer-energy}
\end{equation}
The amplitude $\exp(a)$ carries the stress level, and the shares $\pi_j$ say how much each normalised feature $\tilde g_j$ contributes to the shape, so that a law reads, for instance, as $80\,\%$ of a gradually and $20\,\%$ of a sharply stiffening feature.

For this separation to hold, the features must carry no stress level of their own. We divide each feature by the root-mean-square of its stress over a fixed set of $M_{\mathrm{d}}$ reference states,
\begin{equation}
  \tilde g_j = \frac{g^{(L)}_j}{\omega_j(\bs{\theta})}, \qquad
  \omega_j(\bs{\theta}) = \Big(\frac{1}{M_{\mathrm{d}}} \sum_{n=1}^{M_{\mathrm{d}}} \big\|\bs{S}[g^{(L)}_j](\bs{C}^{\mathrm{d}}_n)\big\|^2\Big)^{1/2},
  \label{eq:featnorm}
\end{equation}
where $\bs{S}[g^{(L)}_j]$ is the stress \eqref{eq:stress} of the PANN with energy $g^{(L)}_j$, corrections \eqref{eq:corrections} included, and $\|\cdot\|$ the Frobenius norm. The reference states are equally spaced stretches $\lambda \in (1, \lambda_{\max}]$ in uniaxial, equibiaxial and plane-strain stretch (Table~\ref{tab:loadcases}), where $\lambda_{\max}$ is a nominal stretch of the material class, e.g.\ $1.3$ for soft tissue and $5$ for elastomers. Neither tested states nor measured stresses enter, so the normalisation is the same for every specimen and test protocol of a material class. The $\omega_j$ are not parameters; they are recomputed from $\bs{\theta}$ in every evaluation, like a normalisation layer. A normalised feature produces stresses of order one over the reference range, and $\exp(a)$ is therefore of the order of the stress of the law over this range, in the units of the data.

The shares follow from unconstrained logits by the softmax function. Since adding the same constant to all logits leaves the shares and the law unchanged, the data carry no information about this direction, and a population over the logits (Section~\ref{sec:hier}) would have a spread along it that describes nothing physical. We remove this non-identifiability by restricting the logits to sum to zero, and write them in coordinates $\bs{z} \in \RR^{N-1}$,
\begin{equation}
  \bs{\pi}(\bs{z}) = \softmax(\bs{V}\bs{z}), \qquad \bs{z} = \bs{V}^\top \log \bs{\pi},
  \label{eq:outlayer}
\end{equation}
where the orthonormal columns of the fixed matrix $\bs{V} \in \RR^{N \times (N-1)}$ span the vectors with zero sum. The coordinates $\bs{z}$ are the isometric log-ratio coordinates of the shares \cite{egozcue2003isometric}; unlike fixing one logit to zero, they treat all features alike. The output layer thus has the $N$ unconstrained parameters $\bs{w} = (a, \bs{z})$, as many as the standard PANN, and the law reads
\begin{equation}
  \bs{S}(\bs{C}; \bs{\theta}, \bs{w}) = \bs{S}\Big[\exp(a) \sum_{j=1}^{N} \pi_j(\bs{z})\, \tilde g_j(\cdot\,; \bs{\theta})\Big](\bs{C}).
  \label{eq:law}
\end{equation}
Each raw feature enters with the weight $\exp(a)\pi_j(\bs{z})/\omega_j(\bs{\theta})$, which takes every positive value, so the representable laws are those of the standard PANN, and every $\bs{w}$ gives a polyconvex law.

The prior of the output layer is
\begin{equation}
  \bs{\pi}(\bs{z}) \sim \mathrm{Dirichlet}\big(\tfrac12 \bs{1}_N\big), \qquad
  a \sim \mathcal{N}\big(\bar a, s_a^2\big),
  \label{eq:single-out}
\end{equation}
which in the coordinates $\bs{z}$ has the density $p(\bs{z}) \propto \prod_{j} \pi_j(\bs{z})^{1/2}$. Compared with the uniform distribution on the shares, the concentration $\tfrac12$ places more probability on compositions dominated by a few features, reflecting the expectation that features specialise and a law combines the few it needs. The log amplitude is vague: $\bar a$ and $s_a$ are the mean and standard deviation of $a$ when the stress level is log-uniform between a lower bound $S_{\min}$ and an upper bound $S_{\max}$, i.e.\ $\bar a = \tfrac12 \log(S_{\min} S_{\max})$ and $s_a = \log(S_{\max}/S_{\min})/\sqrt{12}$. The bounds only have to enclose every plausible stress level of the material class, e.g.\ $1$\,Pa to $1$\,GPa, since the data determine $a$ well; we state them with each data set.

\subsubsection{Posterior}

By Bayes' rule \eqref{eq:bayes}, the posterior of one specimen with $\bs{w} = (a, \bs{z})$ is
\begin{equation}
  p(\bs{\theta}, \bs{w} \mid \mathcal{D}) \propto \prod_{i=1}^{M} \mathcal{N}\big(\hat{\bs{S}}_i \mid \bs{S}(\bs{C}_i; \bs{\theta}, \bs{w}), \sigma_\epsilon^2\big)\; p(\bs{\theta})\; p(\bs{w}),
  \label{eq:post-single}
\end{equation}
with $p(\bs{\theta})$ from \eqref{eq:single} and $p(\bs{w})$ from \eqref{eq:single-out}. When the noise level is inferred, $\sigma_\epsilon$ and its prior join the product.

Every posterior draw is a polyconvex law, and the stresses of the draws in an untested state $\bs{C}_*$ give the uncertainty about the law there. A new measurement $\hat{\bs{S}}_*$ in this state also carries noise and has the posterior predictive distribution
\begin{equation}
  p(\hat{\bs{S}}_* \mid \bs{C}_*, \mathcal{D}) = \int \mathcal{N}\big(\hat{\bs{S}}_* \mid \bs{S}(\bs{C}_*; \bs{\theta}, \bs{w}), \sigma_\epsilon^2\big)\, p(\bs{\theta}, \bs{w} \mid \mathcal{D})\, \diff{\bs{\theta}}\, \diff{\bs{w}}.
  \label{eq:postpred}
\end{equation}
We summarise either distribution by its pointwise quantiles over a range of deformation states and call the region between the $2.5\,\%$ and the $97.5\,\%$ quantile its $95\,\%$ band. The band of the predictive distribution \eqref{eq:postpred} is wider than that of the law by the noise of a new measurement.

\subsection{The hierarchical Bayesian PANN}
\label{sec:hier}

We now consider $K$ specimens of one material, such as aortas of different donors or rubber of different batches, with data $\mathcal{D}_k = \{(\bs{C}_{ik}, \hat{\bs{S}}_{ik})\}_{i=1}^{M_k}$, $k = 1, \dots, K$, and $\mathcal{D} = \{\mathcal{D}_k\}_{k=1}^{K}$. Each specimen follows its own law, and these laws scatter around a mean law. We describe this by a nonlinear mixed-effects model \cite{lindstrom1990nonlinear,gelman2006data} with the PANN as its nonlinear function,
\begin{subequations}
\label{eq:population}
\begin{align}
  \hat{\bs{S}}_{ik} &\sim \mathcal{N}\big(\bs{S}(\bs{C}_{ik}; \bs{\theta}, \bs{w}_k), \sigma_\epsilon^2\big), && i = 1, \dots, M_k, \label{eq:pop-lik} \\
  \bs{w}_k &\sim \mathcal{N}(\bs{\mu}, \bs{\Sigma}), && k = 1, \dots, K. \label{eq:pop-z}
\end{align}
\end{subequations}
All specimens share the hidden layers $\bs{\theta}$, i.e.\ the features, and each has its own output-layer parameters $\bs{w}_k = (a_k, \bs{z}_k)$, the random effects. The mean $\bs{\mu} = (\mu_a, \bs{\mu}_z)$ describes the mean law of the material, and the covariance $\bs{\Sigma}$ how its specimens differ.

The variability sits on the output layer alone, for three reasons. A population over all weights would have hundreds of dimensions, which a handful of specimens cannot determine; on the output layer it has $N$. Every $\bs{w}_k$ gives a polyconvex law, so every specimen drawn from the population is admissible. And the coordinates have a mechanical meaning: a uniformly stiffer specimen differs in its amplitude $a_k$, a specimen that stiffens earlier or later in its shares $\bs{\pi}(\bs{z}_k)$. Sharing learned features and adapting only the last layer follows \cite{snoek2015scalable,harrison2018meta}.

\paragraph{Covariance} A full $N \times N$ covariance is poorly determined by a few specimens. We use
\begin{equation}
  \bs{\Sigma} = \sum_{l=1}^{r} \gamma_l^2\, \bs{u}_l \bs{u}_l^\top + \mathrm{diag}\big(\tau_a^2, \tau_z^2, \dots, \tau_z^2\big),
  \label{eq:pop-sigma}
\end{equation}
which has a direct physical reading. Each unit vector $\bs{u}_l$ is a mode of variation, i.e.\ a direction in which the amplitude and the shares of the specimens vary together, much like the correlated stiffness and stiffening parameters of a classical exponential law. The spread $\gamma_l$ is in log units: one standard deviation along a pure amplitude mode scales the stress of a specimen by $\exp(\gamma_l)$, so that $\gamma_l = 0.3$ means about $\pm 35\,\%$. The residual scales $\tau_a$ and $\tau_z$ collect the scatter of the amplitude and of the shares that no mode explains, and keep $\bs{\Sigma}$ positive definite. The modes $\bs{u}_l$ are point estimates, while the spreads and residual scales carry posterior uncertainty, since they are what a handful of specimens determines least. Since at most $K-1$ modes can be estimated from $K$ specimens, we allow up to $r = 5$ modes for every $K$ and let their prior switch off those the data do not support. A prior on the modes themselves did not change the predictions in our experiments but made training less stable.

\paragraph{Priors} The hidden layers and the mean law receive the priors \eqref{eq:single} and \eqref{eq:single-out} of one specimen. The spreads receive penalised-complexity priors \cite{simpson2017penalising},
\begin{equation}
  \gamma_l \sim \mathrm{Exp}\Big(\frac{-\log\alpha_0}{\gamma_0}\Big), \quad l = 1, \dots, r, \qquad
  \tau_a, \tau_z \sim \mathrm{Exp}\Big(\frac{-\log\alpha_0}{\tau_0}\Big),
  \label{eq:priors}
\end{equation}
which treat a population without variation along a mode, $\gamma_l = 0$, as the base model. They penalise departures from it at a constant rate and have their mode at zero, so that a mode of variation that the data do not support adds no spread to the predictions. Since each mode has its own spread, the modes are switched off individually, as in automatic relevance determination. The rates are set by the probability $\alpha_0 = 0.05$ of exceeding $\gamma_0 = 1$ and $\tau_0 = 0.3$. Physically, specimens of one material class are a priori unlikely to differ in stiffness by more than a factor $e \approx 2.7$ per standard deviation along one mode, and by more than a factor $1.35$ in the residual scatter. Both values can be raised by the same rule for materials whose specimens differ more. The residual scales are not switched off, so that $\bs{\Sigma}$ stays positive definite, and we add a floor of $10^{-3}$ to both. Gamma priors on the precisions \cite{tipping2001sparse} and half-Cauchy priors \cite{gelman2006prior} did not improve our results.

\begin{figure}[htbp]
\centering
\begin{tikzpicture}[scale=0.72, every node/.append style={minimum size=8.5mm}]
  \node[latent] (am)  at (0,0)       {$\mu_a$};
  \node[latent] (nm)  at (2.2,0)     {$\bs{\mu}_z$};
  \node[latent] (sl)  at (4.8,0)     {$\gamma_l$};
  \node[point]  (ul)  at (6.8,0)     {$\bs{u}_l$};
  \node[latent] (ta)  at (9.4,0)     {$\tau_a$};
  \node[latent] (tn)  at (11.2,0)    {$\tau_z$};
  \node[det]    (m)   at (1.1,-2.2)  {$\bs{\mu}$};
  \node[det]    (S)   at (7.6,-2.2)  {$\bs{\Sigma}$};
  \node[latent] (eta) at (4.3,-4.6)  {$\bs{w}_k$};
  \node[obs]    (x)   at (2.0,-7.2)  {$\bs{C}_{ik}$};
  \node[obs]    (y)   at (4.3,-7.2)  {$\hat{\bs{S}}_{ik}$};
  \node[latent] (th)  at (9.6,-7.2)  {$\bs{\theta}$};
  \node[const]  (sy)  at (9.6,-5.2)  {$\sigma_\epsilon$};
  \begin{scope}[on background layer]
    \path (sl.north) ++(0,0.75) coordinate (lt);
    \node[plate, fit=(sl)(ul)(lt), inner sep=3mm, label={[anchor=north west, inner sep=3pt]north west:{\small $l = 1,\dots,r$}}] (pl) {};
    \node[plate, fit=(x)(y), inner xsep=4mm, inner ysep=6.5mm, label={[anchor=south east, inner sep=3pt]south east:{\small $i = 1,\dots,M_k$}}] (pin) {};
    \path (pin.south) ++(0,-0.75) coordinate (kb);
    \node[plate, fit=(eta)(pin)(kb), inner xsep=6mm, inner ysep=3.5mm, label={[anchor=south west, inner sep=3pt]south west:{\small $k = 1,\dots,K$}}] (pk) {};
  \end{scope}
  \draw[edge] (am) -- (m);
  \draw[edge] (nm) -- (m);
  \draw[edge] (sl) -- (S);
  \draw[edge] (ul) -- (S);
  \draw[edge] (ta) -- (S);
  \draw[edge] (tn) -- (S);
  \draw[edge] (m) -- (eta);
  \draw[edge] (S) -- (eta);
  \draw[edge] (eta) -- (y);
  \draw[edge] (x) -- (y);
  \draw[edge] (th) -- (y);
  \draw[edge] (sy) -- (y);
\end{tikzpicture}
\caption{Plate diagram of the hierarchical model \eqref{eq:population}--\eqref{eq:priors}. The hidden-layer parameters $\bs{\theta}$ and the noise level $\sigma_\epsilon$ are shared by all specimens, the parameters $\bs{w}_k$ are drawn once per specimen from the population with mean $\bs{\mu}$ and covariance $\bs{\Sigma}$, and all measurements of a specimen share its $\bs{w}_k$. Circles are latent variables, double circles deterministic functions of their parents and shaded circles observations. The shaded rectangle is a known constant, which becomes latent when the noise level is inferred, the open rectangle is a point estimate, and the dashed rectangles are plates over the indicated index.}
\label{fig:plate}
\end{figure}
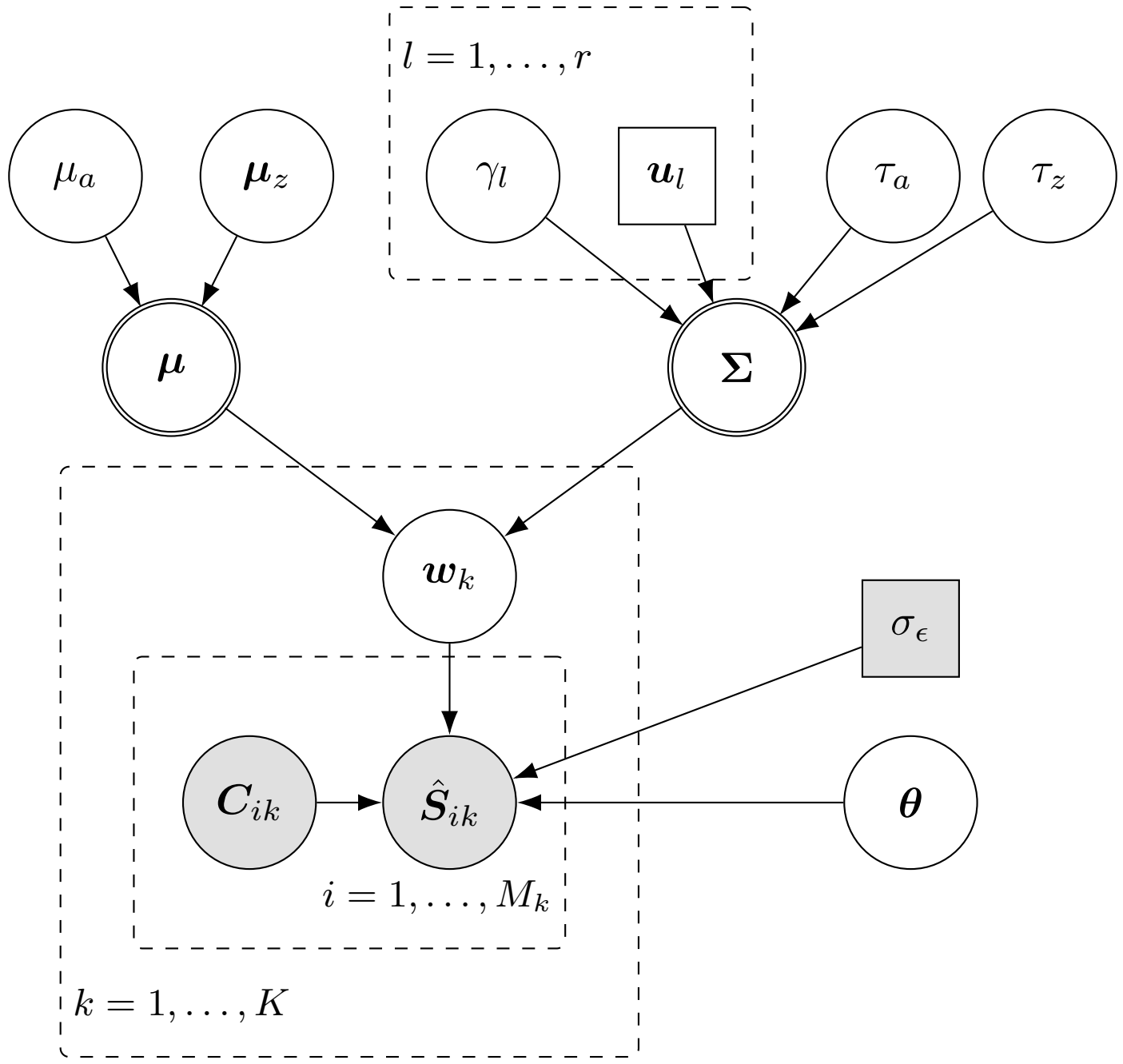

\paragraph{Posterior} The posterior of the hierarchical model is
\begin{equation}
  \begin{split}
  p(\bs{\theta}, \bs{\mu}, \bs{\gamma}, \bs{\tau}, \bs{w}_{1:K} \mid \mathcal{D}) \propto{}& \prod_{k=1}^{K}\Big[\prod_{i=1}^{M_k} \mathcal{N}\big(\hat{\bs{S}}_{ik} \mid \bs{S}(\bs{C}_{ik}; \bs{\theta}, \bs{w}_k), \sigma_\epsilon^2\big)\; \mathcal{N}(\bs{w}_k \mid \bs{\mu}, \bs{\Sigma})\Big]\\
  &\times p(\bs{\theta})\, p(\bs{\mu})\, p(\bs{\gamma})\, p(\bs{\tau}),
  \end{split}
  \label{eq:joint}
\end{equation}
shown as a plate diagram in Figure~\ref{fig:plate}. The Gaussian population is an assumption suited to few specimens. Since it acts on learned features, the resulting population of stress responses is not Gaussian, but skewed or multimodal populations require richer families such as diffusion models \cite{tac2024generative}, which in turn require more specimens.

\subsection{Inference}
\label{sec:inference}

Neither posterior, \eqref{eq:post-single} nor \eqref{eq:joint}, is available in closed form, since the stress is nonlinear in the network parameters. We approximate each by a Gaussian $q$ in the unconstrained unknowns, with the scales $\gamma_l$, $\tau_a$, $\tau_z$ and, if inferred, $\sigma_\epsilon$ represented by their logarithms. We fit $q$ by maximising the evidence lower bound \cite{hoffman2013stochastic,blei2017variational}
\begin{equation}
  \begin{split}
  \mathcal{L}(q) &= \E_{q}\big[\log p(\mathcal{D}, \text{unknowns})\big] + \mathbb{H}[q]\\
  &= \log p(\mathcal{D}) - \KL\big(q \,\|\, p(\text{unknowns} \mid \mathcal{D})\big),
  \end{split}
  \label{eq:elbo-general}
\end{equation}
where $p(\mathcal{D}, \text{unknowns})$ is the product of likelihood and prior, i.e.\ the unnormalised posterior \eqref{eq:post-single} or \eqref{eq:joint}, and $\mathbb{H}$ the entropy. The first term rewards laws that fit the data and are plausible under the prior; the entropy keeps $q$ from collapsing to a single law. We estimate the gradient from reparameterised draws of $q$ \cite{kingma2014autoencoding} and maximise with Adam \cite{kingma2014adam}; the settings are listed in Table~\ref{tab:setup}.

\paragraph{Approximating family} For one specimen, $q$ is a Gaussian with full covariance over all $\mathcal{O}(10^{2})$ unknowns,
\begin{equation}
  q(\bs{\theta}, \bs{w}) = \mathcal{N}\big(\bs{m}, \bs{L}\bs{L}^\top\big),
  \label{eq:qsingle}
\end{equation}
which captures the correlations between all parameters. For the hierarchical model,
\begin{equation}
  q = q_0(\bs{\theta}, \bs{\mu}, \bs{\gamma}, \bs{\tau}) \prod_{k=1}^{K} q_k(\bs{w}_k),
  \label{eq:q}
\end{equation}
where $q_0$, over the unknowns shared by all specimens, is a Gaussian whose covariance is of rank $10$ plus diagonal, and each $q_k$ is a Gaussian with full $N \times N$ covariance, which captures how the amplitude and the shares of one specimen trade off against each other in the fit. The specimens are parameterised directly by $\bs{w}_k$ rather than by their offset from $\bs{\mu}$, which suits densely measured specimens \cite{papaspiliopoulos2007general}. The modes $\bs{u}_l$ are optimised jointly with $q$. The ELBO of the hierarchical model in \eqref{eq:elbo-general} with the posterior  of \eqref{eq:joint}, is
\begin{equation}
  \begin{split}
  \mathcal{L} = {}&\E_q\Big[\sum_{k=1}^{K}\Big(\log p(\mathcal{D}_k \mid \bs{\theta}, \bs{w}_k) + \log\mathcal{N}(\bs{w}_k \mid \bs{\mu}, \bs{\Sigma})\Big) + \log p(\bs{\theta}, \bs{\mu}, \bs{\gamma}, \bs{\tau})\Big]\\
  &+ \mathbb{H}[q_0] + \sum_{k=1}^{K}\mathbb{H}[q_k].
  \end{split}
  \label{eq:elbo}
\end{equation}
It  consists of the data fit of every specimen, the population term $\log\mathcal{N}(\bs{w}_k \mid \bs{\mu}, \bs{\Sigma})$ that ties the specimens together, the priors, and the entropy of $q_0$ and $q_k$.

\paragraph{Initialisation} The variational fit starts from maximum a posteriori fits. For one specimen, we take the best of five, by their objective, as the mean of $q$, with an initial standard deviation of $0.05$. For the hierarchical model, a start from one law fitted to all specimens would offer features that cannot express the differences between them. We therefore take the best of eight fits with shared hidden layers, specimen-specific output layers and no population term, started from the pooled law and from prior draws whose units have their kinks at randomly chosen training states (Algorithm~\ref{alg:hier}). At a known noise level above $1\,\%$ of the stress scale, these fits first weight the data as if the noise were $1\,\%$ and relax to the actual level, which keeps them from collapsing onto one shared law; this affects only the start, not the posterior. We run one variational fit from this start and do not select among several.

\begin{algorithm}[t]
\caption{Variational inference for the hierarchical model.}
\label{alg:hier}
\begin{algorithmic}[1]
\Require data $\mathcal{D}_{1:K}$, noise level $\sigma_\epsilon$ (or its prior), rank $r$
\State standardise the invariants \eqref{eq:standardise}; fix the reference states of \eqref{eq:featnorm}
\State initialise $q$ and $\bs{u}_{1:r}$ from the best of eight MAP fits without population term
\Repeat
  \State draw the shared unknowns from $q_0$ and the specimen parameters from $q_1, \dots, q_K$
  \State evaluate the stresses $\bs{S}(\bs{C}_{ik}; \bs{\theta}, \bs{w}_k)$ via \eqref{eq:featnorm}, \eqref{eq:law}
  \State estimate the ELBO \eqref{eq:elbo} and its gradient; update $q$ and $\bs{u}_{1:r}$ with Adam
\Until{the prescribed number of steps}
\State \Return $q$, $\bs{u}_{1:r}$
\end{algorithmic}
\end{algorithm}

\subsection{Prediction}
\label{sec:prediction}

The fitted population is used for specimens that were not part of the fit. Two kinds of unknowns enter: the shared ones, $(\bs{\theta}, \bs{\mu}, \bs{\gamma}, \bs{\tau})$, which the training data $\mathcal{D}$ determine through the posterior \eqref{eq:joint}, and the parameters $\bs{w}_*$ of the new specimen, which are unknown since the specimen is not part of the training data. The shared unknowns are drawn from their posterior. The parameters $\bs{w}_*$ are drawn from the population if the new specimen has no data, and from their posterior given its data if it has. We state the densities for the three use cases below. In all of them, the approximation $q_0$ of Section~\ref{sec:inference} takes the place of the marginal $p(\bs{\theta}, \bs{\mu}, \bs{\gamma}, \bs{\tau} \mid \mathcal{D})$ of \eqref{eq:joint}.

\subsubsection{New specimen without data}

For a new specimen without data, the parameters $\bs{w}_*$ and the shared unknowns have the joint posterior predictive density
\begin{equation}
  p(\bs{\theta}, \bs{\mu}, \bs{\gamma}, \bs{\tau}, \bs{w}_* \mid \mathcal{D}) = \mathcal{N}(\bs{w}_* \mid \bs{\mu}, \bs{\Sigma})\; p(\bs{\theta}, \bs{\mu}, \bs{\gamma}, \bs{\tau} \mid \mathcal{D}),
  \label{eq:newspec}
\end{equation}
i.e.\ its parameters are a draw from the population, and the population a draw from its posterior. The first factor describes the variability between specimens, the second what the finite number of specimens leaves open about the population. Every draw is an ordinary PANN with fixed parameters, with $\omega_j(\bs{\theta})$ evaluated at the reference states. At every deformation state, the draws give a distribution of the stress of a new specimen, whose $5\,\%$ and $95\,\%$ quantiles, for instance, enclose that stress with $90\,\%$ probability. The draws can also enter a finite element simulation like any hyperelastic law, which gives the distribution of a structural quantity, such as the maximal stress in a component made from a new batch.

\subsubsection{New specimen from a partial test} \label{sec:partial-test}

For a new batch or patient, a short test $\mathcal{D}_* = \{(\bs{C}_i, \hat{\bs{S}}_i)\}_{i=1}^{M_*}$, e.g.\ uniaxial tension up to a small stretch, is often feasible where a full campaign is not. The test turns the prior of $\bs{w}_*$ in \eqref{eq:newspec} into its posterior, and the population then predicts the specimen beyond the tested range and in untested load cases. Whether the test should also update the shared unknowns is a modelling choice, which leads to two densities. The full posterior
\begin{equation}
  \begin{split}
  p(\bs{\theta}, \bs{\mu}, \bs{\gamma}, \bs{\tau}, \bs{w}_* \mid \mathcal{D}, \mathcal{D}_*) \propto{}& p(\mathcal{D}_* \mid \bs{\theta}, \bs{w}_*)\; \mathcal{N}(\bs{w}_* \mid \bs{\mu}, \bs{\Sigma})\\
  &\times p(\bs{\theta}, \bs{\mu}, \bs{\gamma}, \bs{\tau} \mid \mathcal{D})
  \end{split}
  \label{eq:ns-full}
\end{equation}
is the density \eqref{eq:newspec} multiplied by the likelihood of the test; it is the posterior of the hierarchical model with the new specimen as specimen $K+1$, the training specimens integrated out. The cut posterior keeps the shared unknowns at their posterior given $\mathcal{D}$ and conditions only the new specimen on its test,
\begin{equation}
  \begin{gathered}
  p(\bs{\theta}, \bs{\mu}, \bs{\gamma}, \bs{\tau} \mid \mathcal{D})\; p(\bs{w}_* \mid \mathcal{D}_*, \bs{\theta}, \bs{\mu}, \bs{\gamma}, \bs{\tau}), \\
  p(\bs{w}_* \mid \mathcal{D}_*, \bs{\theta}, \bs{\mu}, \bs{\gamma}, \bs{\tau}) = \frac{p(\mathcal{D}_* \mid \bs{\theta}, \bs{w}_*)\, \mathcal{N}(\bs{w}_* \mid \bs{\mu}, \bs{\Sigma})}{p(\mathcal{D}_* \mid \bs{\theta}, \bs{\mu}, \bs{\gamma}, \bs{\tau})}.
  \end{gathered}
  \label{eq:sir}
\end{equation}
The two differ only by the factor $p(\mathcal{D}_* \mid \bs{\theta}, \bs{\mu}, \bs{\gamma}, \bs{\tau}) = \int p(\mathcal{D}_* \mid \bs{\theta}, \bs{w}_*)\, \mathcal{N}(\bs{w}_* \mid \bs{\mu}, \bs{\Sigma})\, \mathrm{d}\bs{w}_*$, the probability of the test given the shared unknowns, with which the full posterior reweights them. The cut posterior thus gives the prediction for a new specimen at almost no cost, since it requires no new fit. The test, however, does not update the features and the population: these remain those learned from the $K$ training specimens, whereas under the full posterior they also reflect the new specimen, which may matter when $K$ is small. Under either density, the stresses $\bs{S}(\bs{C}; \bs{\theta}, \bs{w}_*)$ at untested states give the prediction, summarised by pointwise quantiles as above. The difference to a new specimen without data lies in $\bs{w}_*$: without data it ranges over the whole population, so that its prediction covers the variability of all specimens; with a test, it is confined to the parameters compatible with that test, and the prediction narrows from the population to the specimen. Without a test, both densities reduce to \eqref{eq:newspec}. The experiments use Algorithm~\ref{alg:sir}, which proposes the new specimen from the population and reweights the joint draws by the likelihood of the test, and thereby approximates the full posterior \eqref{eq:ns-full}.

\begin{algorithm}[htbp]
\caption{Prediction of a new specimen from a partial test, as implemented.}
\label{alg:sir}
\begin{algorithmic}[1]
\Require partial test $\mathcal{D}_*$, $N_{\mathrm{pred}}$ draws $(\bs{\theta}^{(s)}, \bs{\mu}^{(s)}, \bs{\gamma}^{(s)}, \bs{\tau}^{(s)}, \bs{w}_*^{(s)})$ of \eqref{eq:newspec}
\State weight each draw by $p(\mathcal{D}_* \mid \bs{\theta}^{(s)}, \bs{w}_*^{(s)})$ and normalise; compute the effective sample size
\State resample $N_{\mathrm{pred}}$ draws with these weights, an approximation of \eqref{eq:ns-full}
\State \Return the stresses of the resampled draws at the states of interest and their quantiles
\end{algorithmic}
\end{algorithm}

\subsubsection{Cohort assignment}
\label{sec:cohort}

Specimens of one material class often fall into cohorts, such as healthy and aneurysmal tissue. With labeled specimens, all cohorts are fitted in one hierarchical model with shared hidden layers: cohort $c$ has its own mean $\bs{\mu}^{[c]}$, spreads, residual scales and modes, and specimen $k$ of cohort $c_k$ follows $\bs{w}_k \sim \mathcal{N}(\bs{\mu}^{[c_k]}, \bs{\Sigma}^{[c_k]})$ in place of \eqref{eq:pop-z}. The cohorts thus combine the same features and differ in how they weight them. A new specimen of unknown cohort follows the mixture $\bs{w}_* \sim \sum_{c=1}^{C} P(c)\, \mathcal{N}(\bs{\mu}^{[c]}, \bs{\Sigma}^{[c]})$, with equal prior probabilities $P(c) = 1/C$. The density of interest is the posterior probability of its cohort given its test,
\begin{align}
  P(c \mid \mathcal{D}_*, \mathcal{D}) &= \frac{P(c)\, p(\mathcal{D}_* \mid c, \mathcal{D})}{\sum_{c'} P(c')\, p(\mathcal{D}_* \mid c', \mathcal{D})}, \\
  p(\mathcal{D}_* \mid c, \mathcal{D}) &= \int p(\mathcal{D}_* \mid \bs{\theta}, \bs{w}_*)\, \mathcal{N}(\bs{w}_* \mid \bs{\mu}^{[c]}, \bs{\Sigma}^{[c]}) \nonumber\\
  &\qquad \times p(\bs{\theta}, \bs{\mu}^{[c]}, \bs{\gamma}^{[c]}, \bs{\tau}^{[c]} \mid \mathcal{D})\, \mathrm{d}\bs{w}_*\, \mathrm{d}\bs{\theta}\, \mathrm{d}\bs{\mu}^{[c]}\, \mathrm{d}\bs{\gamma}^{[c]}\, \mathrm{d}\bs{\tau}^{[c]},
  \label{eq:marglik}
\end{align}
where the marginal likelihood $p(\mathcal{D}_* \mid c, \mathcal{D})$ is the normalising constant of the full posterior \eqref{eq:ns-full} under the population of cohort $c$. Two cohorts are compared by the log Bayes factor
\begin{equation}
  \log B_{cc'} = \log p(\mathcal{D}_* \mid c, \mathcal{D}) - \log p(\mathcal{D}_* \mid c', \mathcal{D}),
  \label{eq:bayesfactor}
\end{equation}
which is positive if cohort $c$ explains the test better, with a magnitude that measures the strength of the evidence. Since the model only approximates the true population and is fitted to few specimens, we do not take its Bayes factors at face value, but calibrate them on the labeled specimens before reading them as probabilities. For each labeled specimen $j$, the model is fitted without it, and $j$ receives
the log Bayes factor $\ell_j$ of its test. A logistic map $P(c \mid \ell) = \mathrm{sigmoid}\big(t\,(\ell - b)\big)$,
with scale $t > 0$ and offset $b$, is fitted to the pairs $(\ell_j, c_j)$ under the priors
$\log t \sim \mathcal{N}(\log 0.3, 1)$ and $b \sim \mathcal{N}(0, 10^2)$ \cite{platt1999probabilistic}, and
the probability of a new specimen is its log Bayes factor mapped through this fit and averaged over the posterior of
$(t, b)$.  To predict the untested curves without committing to a cohort, we fit one
population to all specimens without their labels and condition it on the test as in Section~\ref{sec:partial-test}.

\clearpage
\section{Numerical illustrations}
\label{sec:results}

\newcommand{\experiment}[1]{\paragraph{\normalfont\bfseries #1}}
Both models are meant for noisy and scarce data, so beyond a proof of concept we test how they behave at higher noise levels, how much a fit depends on its random initialization and, for a population, how many specimens $K$ it needs. To control the noise level and the number of specimens, and to compare with a ground truth, most experiments use synthetic laws. We compare the Bayesian PANN with the deterministic PANN \cite{linden2023neural}, whose architecture it shares, and the hierarchical Bayesian PANN with the methods that learn populations of constitutive laws \cite{tac2024generative,mcculloch2025discovering,linka2025}, in the following called competitors. Treloar's rubber \cite{treloar1944stress} and aortas \cite{niestrawska2016microstructure} then test whether the behaviour carries over to measurements (Table~\ref{tab:data}).

\begin{table}[htbp]
\centering
\small
\resizebox{\textwidth}{!}{\begin{tabular}{@{}llllll@{}}
\toprule
section & task & data & load cases & material & reference \\
\midrule
\ref{sec:synthetic-single} & one specimen & synthetic: S-shape law \eqref{eq:sshape} & uniaxial and equibiaxial stretch (PS held out) & compressible, isotropic & deterministic PANN \\
\ref{sec:treloar} & one specimen & real: Treloar's rubber \cite{treloar1944stress} & UT, ET (PS held out) & incompressible, isotropic & benchmark of \cite{holthusen2026benchmarking} \\
\ref{sec:competitors} & population & synthetic: incompressible upturn family \eqref{eq:upturninc} & UT, ET, PS & incompressible, isotropic & competitors \\
\ref{sec:newspecimen} & new specimen & synthetic: both upturn families & UT only, partially & both & competitors, no population \\
\ref{sec:aorta} & cohort assignment & real: aortas, 6 healthy and 6 aneurysmal & biaxial: circumferential, axial & incompressible, transversely isotropic & pooled laws (null test) \\
\bottomrule
\end{tabular}}
\caption{Overview of the numerical illustrations of Section~\ref{sec:results}. On the synthetic data we vary the noise level, from $1\,\%$ to $25\,\%$ for one specimen and $1\,\%$ and $5\,\%$ for populations, and the number of specimens, from $K = 3$ to $24$. UT denotes uniaxial tension, ET equibiaxial tension and PS plane strain (Table~\ref{tab:loadcases}).}
\label{tab:data}
\end{table}

\subsection{Common setup}
\label{sec:setup}
\label{sec:setup-data}

All experiments use homogeneous deformations in three load cases, uniaxial tension (UT), equibiaxial tension (ET) and plane strain (PS), with the deformation gradients of Table~\ref{tab:loadcases}. In every load case the stretch $\lambda$ is prescribed and the nominal stress $P_{11}$, the force per undeformed area with $\bs{P} = \bs{F}\bs{S}$, is measured. For the compressible synthetic laws we prescribe the lateral stretches as $\lambda^{-\nu}$ with $\nu = 0.3$ (\ref{app:setup}). For incompressible materials the lateral stretches follow from $J = 1$, and the hydrostatic pressure from the traction-free thickness direction.

\begin{table}[htbp]
\centering
\small
\begin{tabular}{@{}llll@{}}
\toprule
load case & $\bs{F}$, compressible ($\nu = 0.3$) & $\bs{F}$, incompressible & measured \\
\midrule
UT & $\mathrm{diag}(\lambda, \lambda^{-\nu}, \lambda^{-\nu})$ & $\mathrm{diag}(\lambda, \lambda^{-1/2}, \lambda^{-1/2})$ & $P_{11}$ \\
ET & $\mathrm{diag}(\lambda, \lambda, \lambda^{-2\nu})$ & $\mathrm{diag}(\lambda, \lambda, \lambda^{-2})$ & $P_{11} = P_{22}$ \\
PS & $\mathrm{diag}(\lambda, 1, \lambda^{-\nu})$ & $\mathrm{diag}(\lambda, 1, \lambda^{-1})$ & $P_{11}$ \\
\bottomrule
\end{tabular}
\caption{Deformation gradients of the load cases uniaxial tension (UT), equibiaxial tension (ET) and plane strain (PS) as functions of the stretch $\lambda$, with $\bs{C} = \bs{F}^\top\bs{F} = \bs{F}^2$, and the stress component measured in each. For the compressible laws, UT and ET are a uniaxial and an equibiaxial stretch with prescribed lateral stretches. The aorta data are biaxial with two measured components (Section~\ref{sec:aorta}).}
\label{tab:loadcases}
\end{table}

A synthetic measurement is the stress of the ground-truth law perturbed by Gaussian noise,
\begin{equation}
  \hat{P}_{11,i} = P_{11}^{\mathrm{true}}(\lambda_i) + \epsilon_i, \qquad \epsilon_i \sim \mathcal{N}(0, \sigma_\epsilon^2), \qquad \sigma_\epsilon = \rho\, \max_{\lambda} \big|P_{11}^{\mathrm{true}}(\lambda)\big|,
  \label{eq:noise-synth}
\end{equation}
where the relative noise level $\rho$ is set by each experiment and the maximum is taken over the stretch range of each load case separately, so that every load case is measured with the same relative precision. For a population, the maximum is the median over the population of the peak stress, the same for all specimens, and the population figures show the stresses in these units, labelled \emph{scaled}. The noise level is known to the model and fixed in the likelihood \eqref{eq:lik}; for the real data it is inferred.

The settings of the fits and the computation times are listed in \ref{app:setup}, and intervals and scores are defined in~\ref{app:evaluation}. The code, including our adapted versions of the published methods, will be made available upon publication at \url{https://github.com/pkmtum/Hierarchical-Bayesian-PANN}.

\FloatBarrier
\subsection{One specimen}
\label{sec:onespecimen-results}

We first test the Bayesian PANN on a single specimen: whether its band covers the truth in a load case not used for training, whether its fit stays plausible at high noise, and whether fits from different initializations agree.

\subsubsection{Synthetic law: sensitivity to noise and initialization}
\label{sec:synthetic-single}

\begin{table}[t]
\centering
\small
\begin{tabular}{@{}ccccc@{}}
\toprule
$c_1$ & $c_2$ & $a$ & $d$ & $\kappa$ \\
\midrule
$0.20$ & $0.40$ & $2\cdot 10^{-4}$ & $2.0$ & $5/6$ \\
\bottomrule
\end{tabular}
\caption{Parameters of the S-shape law \eqref{eq:sshape} of Experiments~1 and~2. The parameter $d = 2c_1 + 4c_2$ makes the stress vanish in the undeformed state.}
\label{tab:sshape}
\end{table}

The synthetic specimen follows the S-shape law
\begin{equation}
  \Psi = c_1 (I_1 - 3) + c_2 (I_2 - 3) + a\,(I_1 - 3)^3 - d \log J + \tfrac{\kappa}{2}(J-1)^2,
  \label{eq:sshape}
\end{equation}
with the parameters of Table~\ref{tab:sshape}. In uniaxial stretch it stiffens, flattens and stiffens again, which makes it a demanding target among polyconvex laws. We generate $N = 30$ points each in UT ($\lambda \in [0.8, 5]$) and ET ($\lambda \in [0.8, 2.5]$) and perturb them with noise \eqref{eq:noise-synth} of escalating levels $\rho \in \{1, 2, 5, 10, 25\}\,\%$. Both models are trained on UT and ET and tested on PS over $\lambda \in [1, 3]$, where we compare with the noise-free truth.

\begin{figure}[htbp]
\centering
\includegraphics[width=0.97\textwidth]{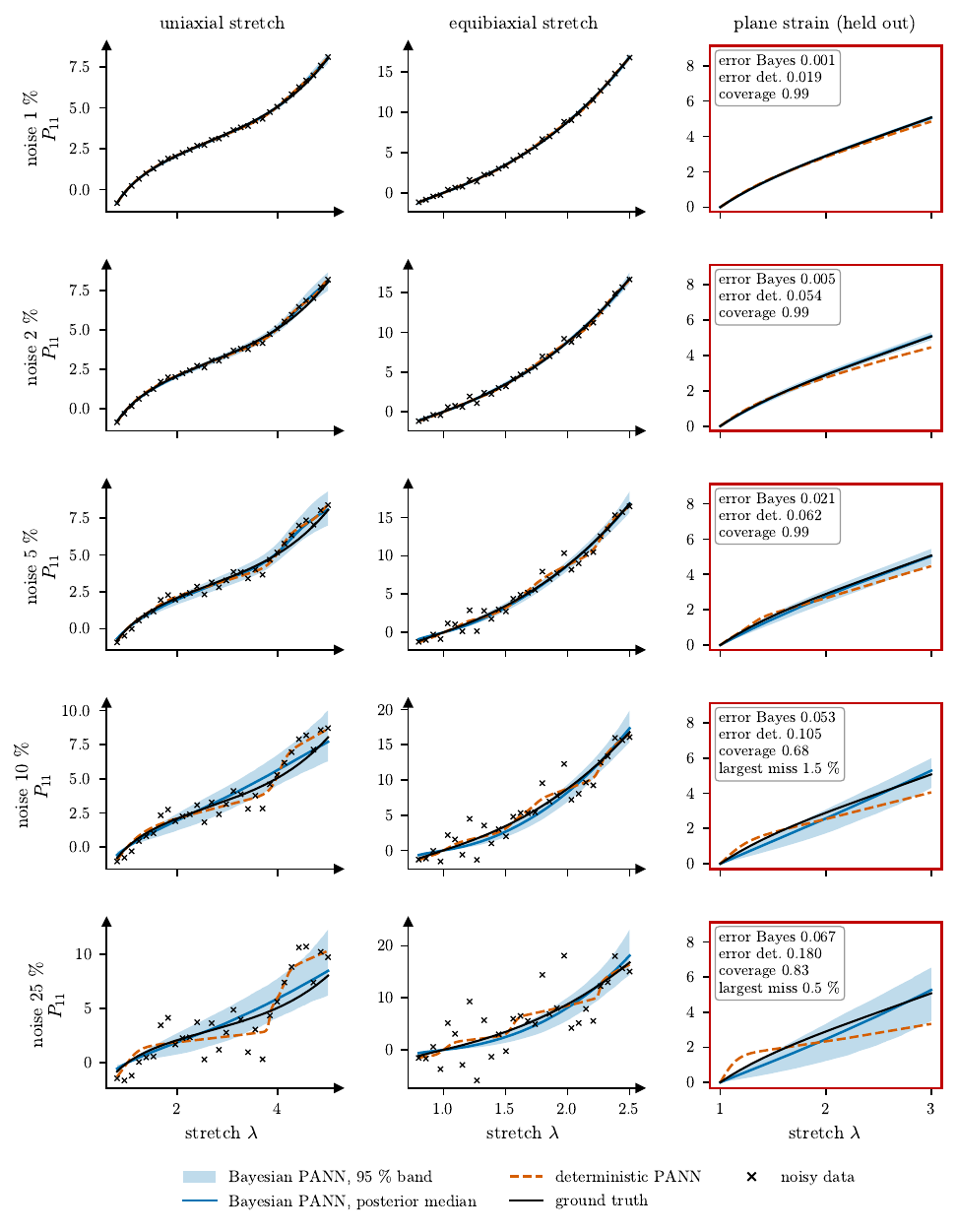}
\caption{Sensitivity of the single-specimen fit to the noise level $\rho$ \eqref{eq:noise-synth} (Experiment~1). One specimen of the S-shape law \eqref{eq:sshape} is fitted in UT and ET and tested in PS (Table~\ref{tab:loadcases}). The Bayesian PANN (Section~\ref{sec:onespecimen}) gives a $95\,\%$ band of the law, which does not include the noise. The deterministic PANN \cite{linden2023neural} gives a single curve without a band, and of its five restarts we show the one with the lowest training loss. Error and coverage are defined in \ref{app:evaluation}, and the largest miss is the largest distance of the truth outside the band relative to the peak stress.}
\label{fig:single}
\end{figure}

\experiment{Experiment 1: noise sensitivity against the deterministic PANN} Figure~\ref{fig:single} compares the Bayesian PANN with the deterministic PANN \cite{linden2023neural} at increasing noise. At every noise level the band covers the training data, and even at $25\,\%$ noise, with few and very noisy points, it gives a smooth and plausible curve. The band widens with the stretch, which is the expected behaviour of hyperelastic laws with uncertain parameters, also of classical ones. In the held-out PS case the band covers the truth at the noise levels from $1\,\%$ to $5\,\%$. At $10\,\%$ and $25\,\%$ the band lies slightly below the truth (Figure~\ref{fig:single}). Data this noisy carry little information, so that the prior dominates the posterior, and in PS the prior favours stiffening curves: of $2000$ draws of the prior predictive, $1.3\,\%$ are as concave as the true curve.

The deterministic PANN tracks the truth at $1\,\%$ noise. At this level there is little noise to overfit, and the band that the Bayesian PANN adds is narrow enough to be negligible for many applications. From $2\,\%$ noise on, the deterministic fit follows the noise. Its curves develop kinks and plateaus in the trained cases (Figure~\ref{fig:single}, $10\,\%$ and $25\,\%$) and move away from the truth in the held-out case, where its error is two to eleven times that of the Bayesian PANN. This is the failure mode of Section~\ref{sec:intro}, which the prior of the Bayesian PANN prevents; the comparison therefore suggests the Bayesian PANN for data with moderate to large noise.

\begin{figure}[htbp]
\centering
\includegraphics[width=\textwidth]{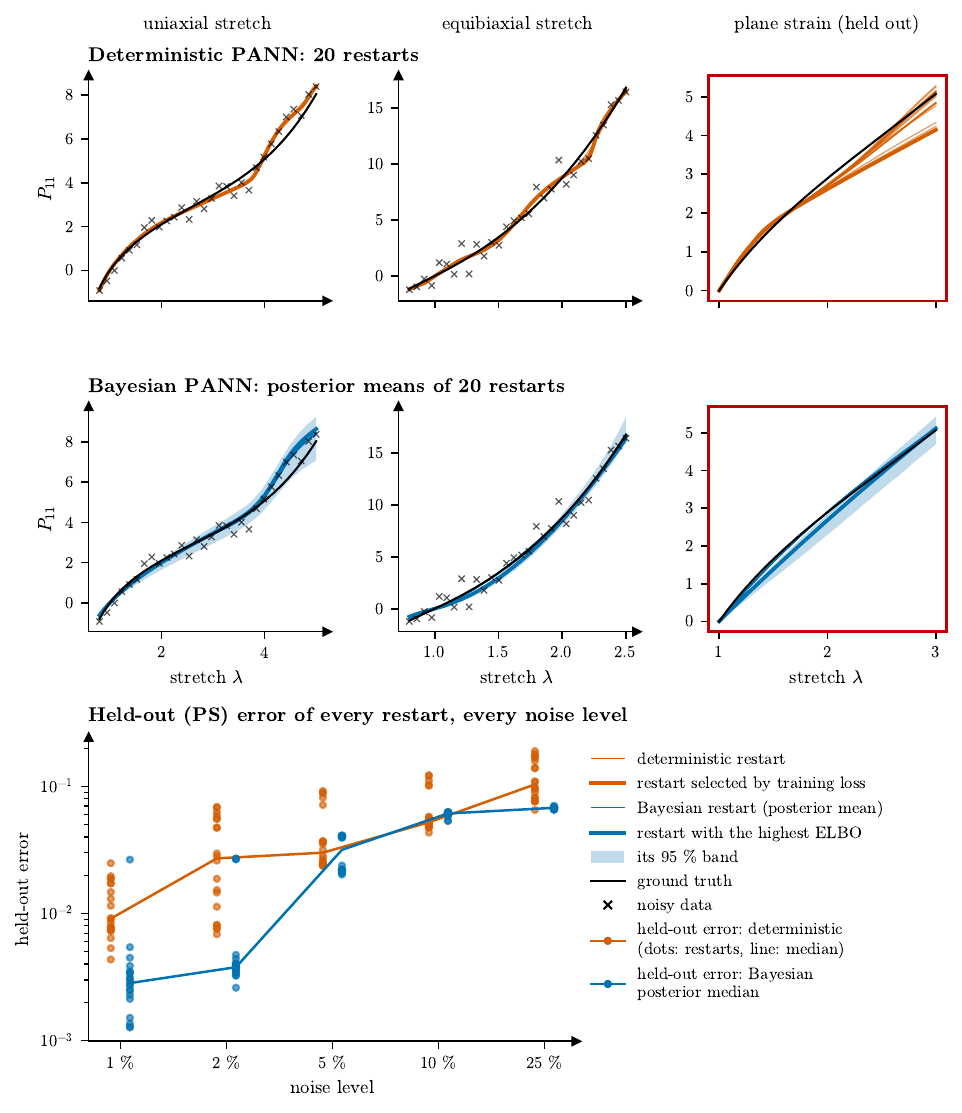}
\caption{Stability of the fits over restarts from different random initializations (Experiment~2). The deterministic PANN \cite{linden2023neural} and the Bayesian PANN (Section~\ref{sec:onespecimen}) are each restarted $20$ times on the data of Figure~\ref{fig:single}, shown at $5\,\%$ noise in the top and middle rows. Each Bayesian restart is drawn as its posterior mean, and only the restart with the highest ELBO is drawn with its band. The bottom panel gives the held-out error (\ref{app:evaluation}) of every restart at every noise level, using the posterior median for the Bayesian PANN. It therefore measures how far the restarts spread and does not reflect the band.}
\label{fig:restarts}
\end{figure}

\experiment{Experiment 2: robustness over restarts} Among the tested methods of a recent benchmark on material law discovery for Treloar's data, the PANN varies most over $100$ initializations \cite{holthusen2026benchmarking}. We therefore repeat Experiment~1 with $20$ restarts of both models at every noise level (Figure~\ref{fig:restarts}). In the trained cases the deterministic restarts nearly coincide, but on a curve that follows the noise rather than the truth. In the held-out case they spread into two groups, visible as two bundles of curves in PS and as two clusters of errors at $5\,\%$ and $10\,\%$ noise, which suggests two local optima of the training loss. The posterior means of the $20$ Bayesian restarts nearly coincide. From $5\,\%$ noise on, the held-out errors of the deterministic restarts spread by a factor of $3$ to $4$, and those of the Bayesian restarts by a factor of $1.1$ to $2$. At $1\,\%$ and $2\,\%$ the deterministic restarts spread by a factor of $6$ and $10$; single Bayesian restarts deviate there as well, but all stay below a held-out error of $0.03$. The error measures the spread of the restarts; it scores a single curve and does not reflect the band. The screening for a negative energy (\ref{app:evaluation}) finds $19$ of the $4\cdot 10^{4}$ posterior draws at $25\,\%$ noise negative, and none at the lower noise levels.

\subsubsection{Treloar's rubber: proof of concept on real data}
\label{sec:treloar}

Treloar's measurements on vulcanised rubber \cite{treloar1944stress} are a standard test of constitutive models and were recently used to benchmark data-driven methods in a common setup \cite{holthusen2026benchmarking}. They contain $14$ points in each of UT ($\lambda \in [1, 7.25]$), ET ($\lambda \in [1, 4.45]$) and PS ($\lambda \in [1, 4.99]$). As for the synthetic law, we train on UT and ET and hold out PS. Since the data carry no ground truth, we score the coverage of the held-out measurements by the predictive band \eqref{eq:postpred}.

\begin{figure}[t]
\centering
\includegraphics[width=\textwidth]{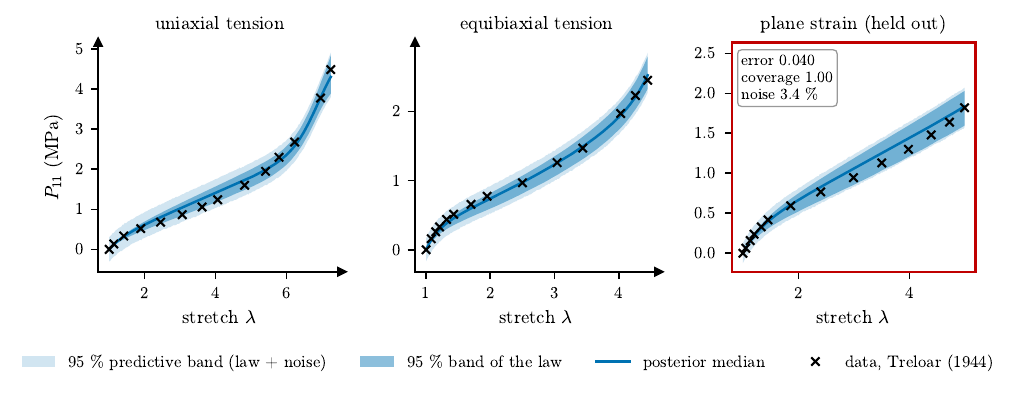}
\caption{Proof of concept of the Bayesian PANN (Section~\ref{sec:onespecimen}) on real data, Treloar's rubber \cite{treloar1944stress}, trained in UT and ET and tested in PS (Table~\ref{tab:loadcases}). The noise level of these data is not known, so it is inferred together with the law, and the inferred level in the box also absorbs model error. The figure shows the restart with the highest ELBO of $20$. Without a ground truth, the error and the coverage (\ref{app:evaluation}) are computed against the measured PS points, and the coverage uses the predictive band, which includes the noise.}
\label{fig:treloar}
\end{figure}

The noise level of these data is not known. A deterministic PANN fitted to them as in the benchmark deviates from the points by about $2\,\%$ of the peak stress, and the most accurate methods of the benchmark do not pass through every point either \cite{holthusen2026benchmarking}. We therefore infer the noise level together with the law and take these $2\,\%$ as $\sigma_{\epsilon,0}$ of its log-normal prior (Section~\ref{sec:onespecimen}). The inferred level contains model error as well as noise, and the predictive band is to be read as such. The posterior places the noise level at $3.4\,\%$, and Figure~\ref{fig:treloar} shows that the $95\,\%$ predictive band then covers every training point and every held-out PS point, in each of the $20$ restarts. Its held-out error is close to the $0.026$ of the deterministic PANN. On these data the Bayesian PANN thus reaches about the accuracy that the PANN architecture allows.

\FloatBarrier
\subsection{Populations: comparison with competing methods}
\label{sec:competitors}

Every downstream task, such as the prediction for a partially tested specimen in Section~\ref{sec:newspecimen}, rests on the population learned from the data. We therefore first test how well the learned population envelops the population of a synthetic law, and compare it with the three competitors, each run with its published code and changed only where our data require it (\ref{app:competitors}): the generative model of Ta\c{c} et al.\ \cite{tac2024generative}, the Gaussian CANN \cite{mcculloch2025discovering} and the Bayesian CANN \cite{linka2025}. The Bayesian CANN fits one law to all specimens and therefore stands for complete pooling. This subsection tests the requirements of calibration: whether the population approaches the true one from few specimens as their number $K$ grows (\ref{req:data}, \ref{req:unc}), whether it is still found at a higher noise level (\ref{req:noise}), and whether it is found more reliably than by the competitors.

\paragraph{Setup} The specimens follow the incompressible upturn family
\begin{equation}
  \Psi_k = 3\left[c_{1,k}(I_1 - 3) + a_k (I_1 - 3)^3\right], \qquad J = 1,
  \label{eq:upturninc}
\end{equation}
whose parameters are drawn for every specimen $k$ from
\begin{align}
  \begin{pmatrix} \log c_{1,k} \\ \log a_k \end{pmatrix} &\sim \mathcal{N}\big(\bs{m}, \bs{\Sigma}_{\mathrm{true}}\big), \qquad
  \bs{m} = \begin{pmatrix} \log 0.2 \\ \log(1.5\cdot 10^{-4}) \end{pmatrix}, \nonumber \\
  \bs{\Sigma}_{\mathrm{true}} &= \begin{pmatrix} 0.25 & 0 \\ 0 & 0.5 \end{pmatrix}
  \begin{pmatrix} 1 & 0.8 \\ 0.8 & 1 \end{pmatrix}
  \begin{pmatrix} 0.25 & 0 \\ 0 & 0.5 \end{pmatrix},
  \label{eq:upturninc-pop}
\end{align}
i.e.\ with standard deviations $0.25$ and $0.5$ (outer factors) and correlation $0.8$ (middle factor). The specimens thus differ in their initial stiffness $c_1$ and in their upturn $a$, and stiffer specimens tend to have a stronger upturn (incompressible, as the competitors are formulated for incompressible materials). Besides the spread of the specimens, a population describes how their properties are related. To check whether a method learns this relation, we compare two properties that can be read off the UT curve of every new specimen, its stiffness and its upturn, by their correlation over new specimens,
\begin{equation}
  r_{\mathrm{su}} = \mathrm{corr}\Big(\underbrace{\log P_{11}(\lambda = 2)}_{\text{stiffness}},\ \underbrace{\log \frac{P_{11}(\lambda = 6)}{P_{11}(\lambda = 3)}}_{\text{upturn}}\Big),
  \label{eq:corr-su}
\end{equation}
which is $+0.46$ in this family. Every specimen is tested in UT ($\lambda \in [0.8, 6]$), ET ($\lambda \in [0.8, 3.5]$) and PS ($\lambda \in [1, 5]$), with $N = 30$ points per load case, at noise levels $\rho = 1\,\%$ and $5\,\%$. We draw four data sets of nested specimens with $K \in \{3, 6, 12, 24\}$, i.e.\ $32$ fits per method. We score every fit by comparing the new specimens of its population with $200$ new specimens drawn from the true population, by the energy distance, the coverage and the band width of~\ref{app:evaluation}.

A comparison with the true population alone would be partly unfair, since $K$ specimens can by chance be unrepresentative of their population. We therefore add a reference, the \emph{known-law ceiling}, which is given the form of the law \eqref{eq:upturninc} and infers only the population \eqref{eq:upturninc-pop} of $(\log c_1, \log a)$ from the same data, with a hierarchical model under weakly informative priors, the prior of the mean centered at the true mean, sampled by NUTS \cite{hoffman2014nuts} (\ref{app:competitors}). It marks the best result that the data allow.

\begin{figure}[htbp]
\centering
\includegraphics[width=0.96\textwidth]{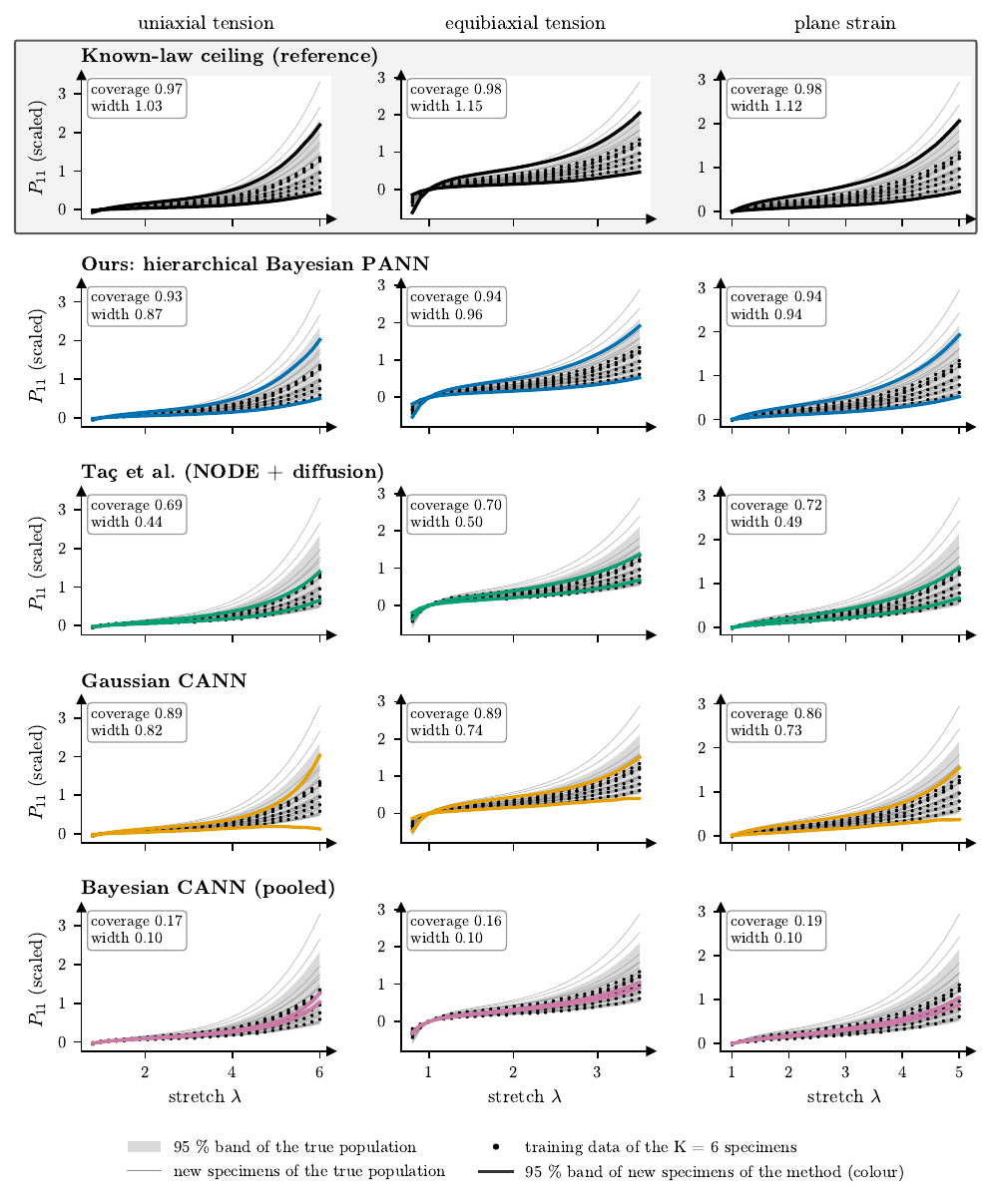}
\caption{Populations learned from few specimens by our model and the competitors (Experiment~3). All methods are fitted to the same $K = 6$ specimens of the incompressible upturn family \eqref{eq:upturninc} at $1\,\%$ noise and compared with the true population \eqref{eq:upturninc-pop}. The known-law ceiling (Section~\ref{sec:competitors}) is given the form of the law and marks the best result that the data allow. Ta\c{c} et al.\ \cite{tac2024generative} and the Gaussian CANN \cite{mcculloch2025discovering} learn a population, whereas the Bayesian CANN \cite{linka2025} fits one law to all specimens, so that its band describes the uncertainty of the mean law rather than the spread of the specimens. Coverage and width (\ref{app:evaluation}) are computed over $200$ new specimens of the true population, and stresses are in the scaled units of Section~\ref{sec:setup-data}.}
\label{fig:comp-bands}
\end{figure}

\experiment{Experiment 3: populations from few specimens} Figure~\ref{fig:comp-bands} shows the populations from $K = 6$ specimens at $1\,\%$ noise. Of the four methods, our population is closest to the true one: in all three load cases its band has about the width and the shape of the true band, and it looks much like the band of the known-law ceiling. The bands of the competitors are visibly narrower than the true band. That of Ta\c{c} et al.\ is the narrowest of the three population methods. That of the Gaussian CANN is closer to the true width but still narrower than ours, and in UT it opens into a funnel at large stretches: its upper edge stiffens sharply towards the end of the range and its lower edge falls below the true band, a shape that the band of the known-law ceiling does not show (\ref{app:competitors}). The Bayesian CANN finds the mean law well, but its band is a thin strip around this law that misses most new specimens, as expected of a pooled fit (Section~\ref{sec:intro-pooling}).

\begin{figure}[htbp]
\centering
\includegraphics[width=\textwidth]{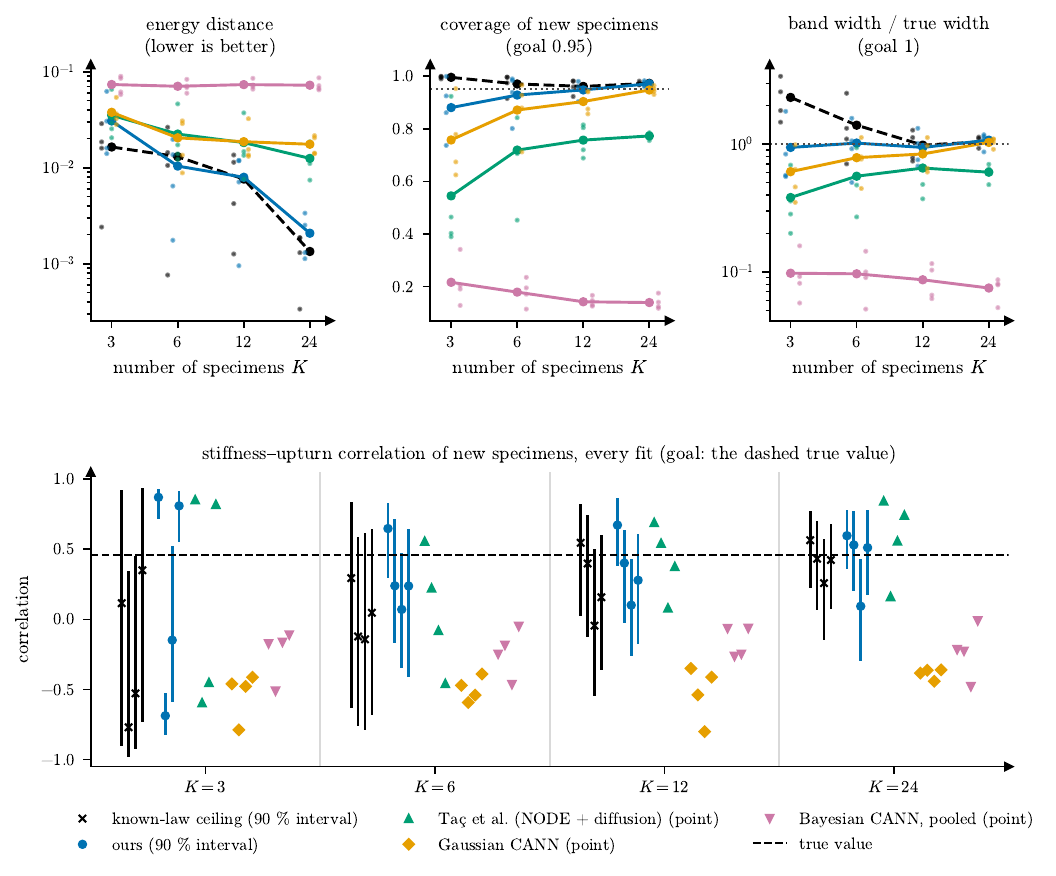}
\caption{Dependence of the learned populations on the number of specimens $K$ (Experiment~3), in the setting of Figure~\ref{fig:comp-bands} with four data sets per $K$. The top row compares the new specimens of each fit with the true population \eqref{eq:upturninc-pop} by the energy distance \eqref{eq:edist}, the coverage and the relative band width (\ref{app:evaluation}), each averaged over the load cases. The bottom row tests whether a population learns how the properties of a specimen are related, by the stiffness--upturn correlation \eqref{eq:corr-su} of its new specimens. Our model and the known-law ceiling quantify their uncertainty about the population and therefore give an interval, whereas the competitors give a point value.}
\label{fig:comp-metrics}
\end{figure}

Figure~\ref{fig:comp-metrics} quantifies the comparison for $K = 3$, $6$, $12$ and $24$ at $1\,\%$ noise. Our population follows the ceiling closely. Its energy distance to the true population decreases with $K$ like that of the ceiling, its bands cover close to the nominal $0.95$ of new specimens and more so as $K$ grows ($0.88$ at $K = 3$, $0.97$ at $K = 24$), and its band width stays close to the true one ($0.94$ to $1.08$ times). The competitors do worse by all three measures up to $K = 12$, i.e.\ when the specimens are few. The two population methods, Ta\c{c} et al.\ and the Gaussian CANN, improve with more specimens, and at $K = 24$ the Gaussian CANN reaches a coverage and a width as close to their targets as ours, at a higher, i.e.\ worse, energy distance. At $K = 3$ the energy distances of the three population methods are similar, but our coverage and width are clearly closer to their targets.

The lower panel of Figure~\ref{fig:comp-metrics} suggests where the difference might come from. It gives the stiffness--upturn correlation \eqref{eq:corr-su} of the new specimens of every fit. Our model returns an interval, since it quantifies its uncertainty about the population, whereas the competitors return a point estimate. Our interval contains the true correlation in $11$ of the $16$ fits, more reliably as $K$ grows, and that of the known-law ceiling in $14$ of its $16$ fits. The point values of Ta\c{c} et al.\ approach the true value as $K$ grows. Those of the Gaussian CANN and the Bayesian CANN are negative in every fit, i.e.\ in their new specimens a stiffer specimen tends to have a smaller upturn, whereas in the true population it tends to have a larger one (\ref{app:competitors}). Compared with the ceiling, our intervals of the correlation are narrower than the spread that the data allow even with the law known, i.e.\ our model is overconfident about how stiffness and upturn are related. This matters most at $K = 3$, the smallest number of specimens for which the covariance of two parameters can be estimated at all. There our interval contains the true correlation in $1$ of $4$ fits, against $2$ of $4$ for the ceiling. The same overconfidence shows in the stress band of new specimens (Figure~\ref{fig:comp-metrics}, upper right). At $K = 3$ the band of the ceiling is $2.3$ times as wide as the $95\,\%$ band of the true population, since three specimens reveal little about the population, whereas ours stays about as wide as the true band. We return to this limitation in Section~\ref{sec:discussion}.

At $5\,\%$ noise our population behaves as at $1\,\%$, whereas the differences to the competitors grow (\ref{app:pop-noise}, Figure~\ref{fig:comp-metrics-5}): the Gaussian CANN, for instance, fails in three of the four values of $K$, with bands more than $30$ times as wide as the true ones on average, since the competitors have no separate place for the noise (\ref{app:competitors}).

\FloatBarrier
\subsection{A new specimen from a partial test}
\label{sec:newspecimen}

We now use the populations of Section~\ref{sec:competitors} to predict a new specimen from a short test of it (Section~\ref{sec:partial-test}), which tests \ref{req:pred}. Here the population has to locate one specimen within it from a few of its points, whereas in Section~\ref{sec:competitors} it had to cover new specimens without any data of them.

\paragraph{Setup} We take the populations for $K = 6$ and $12$, both noise levels and all four data sets. For every data set, $25$ new specimens of the true population are tested in UT up to a stretch $\lambda_{\max} \in \{2, 3, 6\}$, where $\lambda_{\max} = 6$ is the complete UT curve, at the noise level of the population's training data. We condition our population on the test by sampling-importance-resampling \eqref{eq:sir} (Algorithm~\ref{alg:sir}) and predict the remainder of UT, ET and PS. Of the competitors, only Ta\c{c} et al.\ provide a conditioning on new measurements \cite{tac2024generative}, which we use as published. The Gaussian CANN and the Bayesian CANN provide none, so we condition the population that each of them has fitted on the test by Bayes' rule (\ref{app:competitors}). The known-law ceiling is conditioned by sampling-importance-resampling as well. The null reference is a Bayesian PANN fitted to the partial test alone, with the prior of Section~\ref{sec:onespecimen} and without a population.

\begin{figure}[htbp]
\centering
\includegraphics[width=0.88\textwidth]{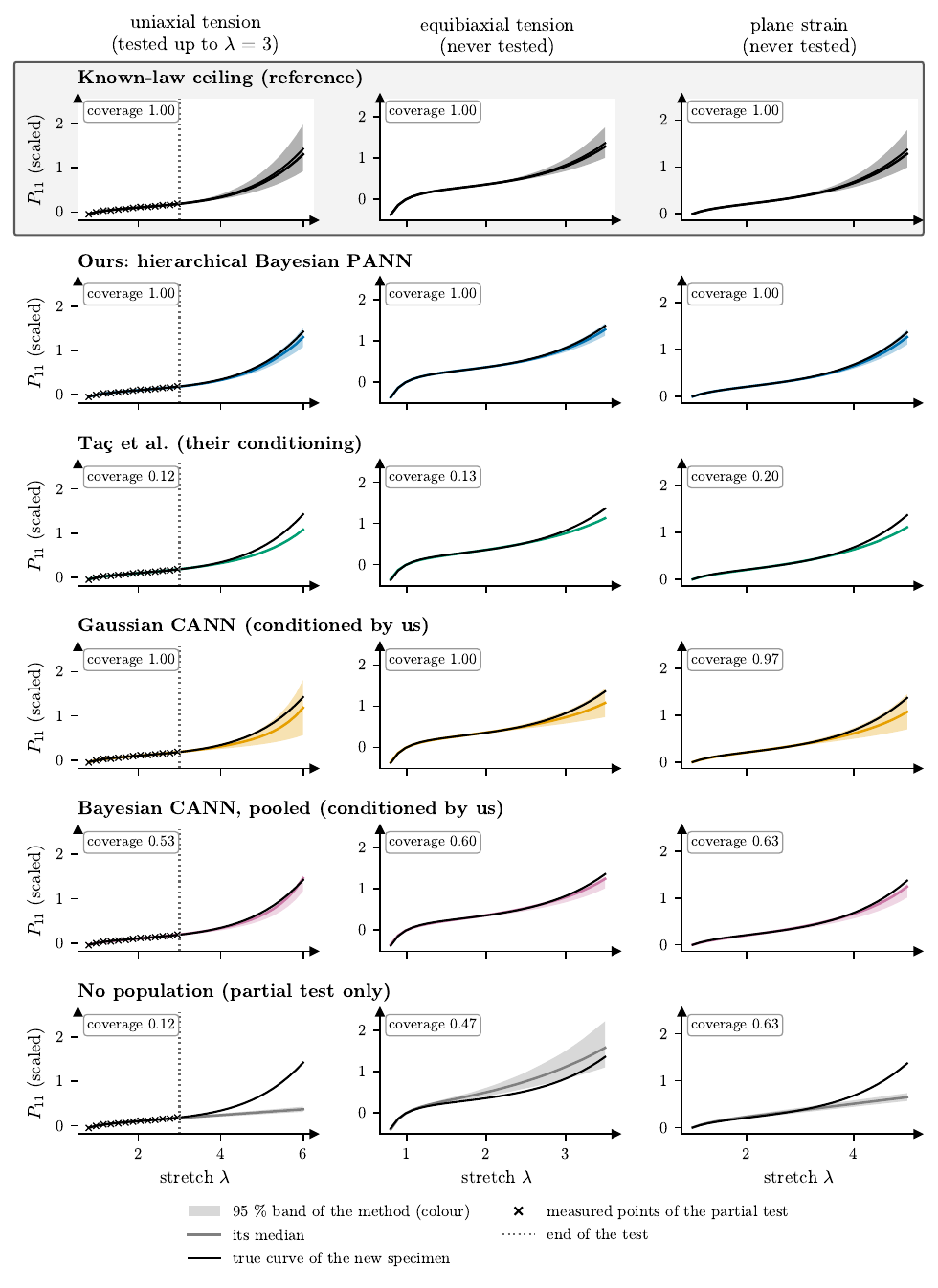}
\caption{Prediction of one new specimen from a partial uniaxial test (Experiment~4). The specimen is tested in UT up to $\lambda = 3$, and the rest of UT, ET and PS is predicted with the populations of Figure~\ref{fig:comp-bands}. We condition our population and the known-law ceiling by sampling-importance-resampling (Algorithm~\ref{alg:sir}) and Ta\c{c} et al.\ \cite{tac2024generative} by their own conditioning. The Gaussian CANN \cite{mcculloch2025discovering} and the Bayesian CANN \cite{linka2025} provide none, so we condition them by Bayes' rule (\ref{app:competitors}). The last row is the Bayesian PANN (Section~\ref{sec:onespecimen}) fitted to the partial test alone. The band of Ta\c{c} et al.\ is narrow enough to look like a single curve. In UT the coverage counts only the part beyond the test, and the specimen is the one on which our model scores at the median of the $25$ new specimens.}
\label{fig:newspecimen}
\end{figure}

\experiment{Experiment 4: prediction from a partial uniaxial test} Figure~\ref{fig:newspecimen} shows one new specimen tested up to $\lambda_{\max} = 3$ at $1\,\%$ noise. Our band covers the specimen's true curve beyond the tested range and in the two untested load cases. Ta\c{c} et al.\ and the fit without a population fail visibly. The conditioned samples of Ta\c{c} et al.\ nearly coincide, so that their band is close to a single curve, which misses the specimen beyond the tested range. The fit without a population follows the data within the tested range. Beyond it and in the untested load cases, it continues the behaviour it has seen and fills the gaps from its prior, which here gives a nearly linear continuation in UT that misses the upturn of the specimen. The two CANNs fail less visibly. The Bayesian CANN returns a narrow band around nearly its pooled law, which happens to lie close to this specimen. The band of the Gaussian CANN spans much of the range of stresses that its population contains, i.e.\ the test hardly narrows it down, but it covers this specimen.

\begin{figure}[htbp]
\centering
\includegraphics[width=\textwidth]{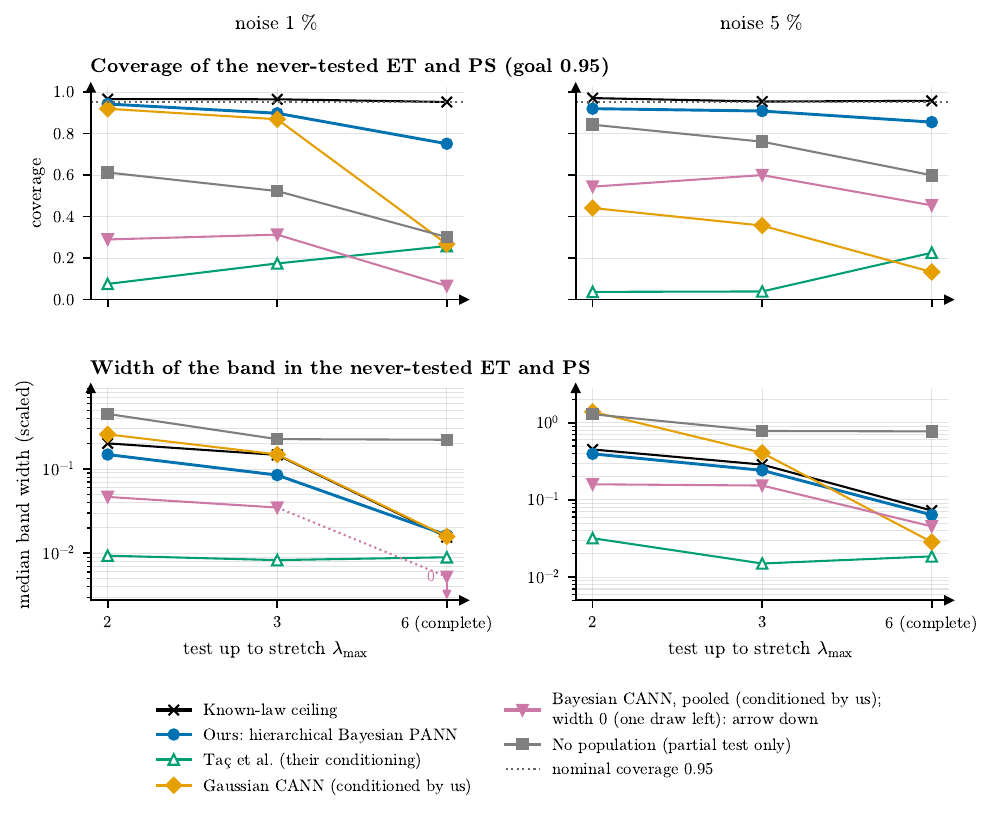}
\caption{Prediction of new specimens from partial uniaxial tests of increasing length (Experiment~4), summarised over all new specimens for populations from $K = 6$ specimens at $1\,\%$ and $5\,\%$ noise. The test runs in UT up to $\lambda_{\max}$, where $\lambda_{\max} = 6$ is the complete UT curve, and the methods are conditioned as in Figure~\ref{fig:newspecimen}. The coverage of the never-tested ET and PS is a mean over the two load cases, over $25$ new specimens per data set and over the data sets, and the band width is a median over the specimens.}
\label{fig:ns-summary}
\end{figure}

Figure~\ref{fig:ns-summary} summarises all specimens, test lengths and both noise levels at $K = 6$, and shows a similar picture as the single specimen of Figure~\ref{fig:newspecimen}. When the UT test ends at $\lambda_{\max} = 2$ or $3$, our band covers $0.90$ to $0.94$ of the true curves of the untested ET and PS at both noise levels, $1\,\%$ and $5\,\%$, close to the $0.95$ to $0.97$ of the ceiling. Ta\c{c} et al.\ cover fewer than $0.2$ of the true curves of ET and PS for these tests, consistent with their narrow population of Section~\ref{sec:competitors}, since conditioning can only select among the laws that a population contains; a looser conditioning width changes this only mildly (\ref{app:competitors}). The Bayesian CANN covers only $0.29$ to $0.31$ of the true curves at $1\,\%$ noise and $0.54$ to $0.60$ at $5\,\%$. The Gaussian CANN is the strongest competitor at $1\,\%$ noise and short tests, where it covers nearly as many of the true curves as ours, but with a band about $1.7$ times as wide. Its prediction is conservative: it carries the general shape of its population over to the new specimen, but it does not learn what the measured part of the curve implies for the rest. At $5\,\%$ noise it covers fewer than half of the true curves for every test length.

A longer test narrows down where the specimen lies within the population, for every method. Since our population is close to the true one, our band in ET and PS narrows around the true curves as the test grows longer (Figure~\ref{fig:ns-summary}, bottom), and it keeps its coverage up to a test to $\lambda_{\max} = 3$. With the complete UT curve at $1\,\%$ noise, however, it becomes overconfident: it narrows to the width of the band of the ceiling, but its coverage of ET and PS falls to $0.75$, against $0.95$ for the ceiling. The effect is weaker with more specimens: averaged over both noise levels, this coverage rises from roughly $0.8$ at $K = 6$ to roughly $0.89$ at $K = 12$ (\ref{app:newspec}, Figure~\ref{fig:ns-summary-k12}). This suggests that the cause is the overconfidence of our population for few specimens (Section~\ref{sec:competitors}), not the conditioning, since the ceiling, conditioned in the same way, keeps its coverage. Our misses concentrate on the specimens that our population considers unlikely, for which few of its draws match the test, whereas the specimens that it represents well are covered at close to $0.9$.

The Gaussian CANN cannot locate a specimen within its population in this way, since its population carries the wrong correlation between stiffness and upturn (Section~\ref{sec:competitors}). A longer test narrows its band about as much as ours, but around biased curves, and with the complete UT curve it covers $0.27$ of the true curves of ET and PS, about a third of our coverage. Even with the complete UT curve, our coverage stays well above that of every competitor and of the fit without a population.

\FloatBarrier
\subsection{Healthy and aneurysmal aortas}
\label{sec:aorta}

Soft tissues are often tested in cohorts, such as healthy and diseased tissue. For a new specimen it is then of interest to which cohort it belongs (\ref{req:cohort}) and how it responds in the tests that were not made (\ref{req:pred}). As a proof of concept on real data, we use biaxial tests of six healthy and six aneurysmal human abdominal aortas \cite{niestrawska2016microstructure}, each loaded in the circumferential and the axial direction of the vessel. In both experiments below, the new specimen is tested only in the circumferential direction.

\paragraph{Data} We use the data of Niestrawska et al.\ \cite{niestrawska2016microstructure} in the form reported by Linka et al.\ \cite{linka2025}. Since they are not openly available, we digitised them from published figures and completed the curves that end before the stiff regime with synthetic points (\ref{app:aortadata}).

\paragraph{Setup} The network is the transversely isotropic incompressible PANN of Linden et al.\ \cite{linden2023neural} with one fibre direction along the circumferential axis (\ref{app:ti}). The noise level is not known and is inferred, one level per population, under the log-normal prior of Section~\ref{sec:onespecimen} with $\sigma_{\epsilon,0} = 1$\,kPa. The stress bounds behind the prior of $\bar a$ are $1$\,Pa and $1$\,GPa. As in Experiment~2, every fit is repeated from three random initializations (\ref{app:setup}). Experiment~5 keeps, for every fit, the restart with the highest ELBO, as for Treloar's rubber (Section~\ref{sec:treloar}); Experiment~6 reports the mean over the three restarts. To score the methods on the twelve available specimens, we hold out each specimen in turn and treat it as new, i.e.\ every model used for it is fitted without it.

The null test is complete pooling, one Bayesian CANN per cohort (Section~\ref{sec:competitors}), which Linka et al.\ \cite{linka2025} fit to the same data (\ref{app:competitors}). It assigns a held-out specimen to the cohort whose law gives the circumferential curve the larger predictive likelihood, and predicts with the law of that cohort. Linka et al.\ compare the cohorts only qualitatively \cite{linka2025}, so this rule is our addition.

\begin{figure}[htbp]
\centering
\includegraphics[width=\textwidth]{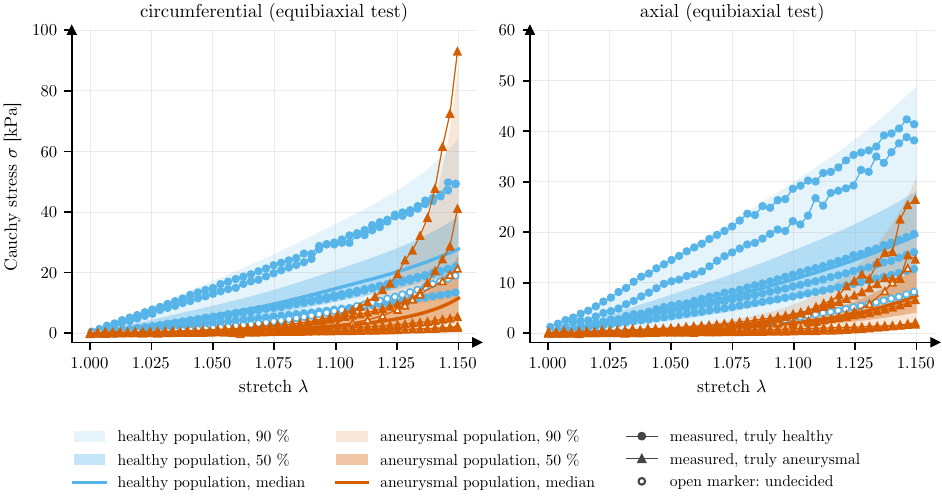}
\caption{Cohort populations of the aortas, for illustration (Experiment~5). The populations come from a joint fit (Section~\ref{sec:cohort}) to all twelve specimens with their labels, which enters neither the assignment of Figure~\ref{fig:aorta-pcal} nor the prediction of Figure~\ref{fig:predict-unl}. The bands are those of new specimens of each population without the noise, from the restart with the highest ELBO. Open markers are the two specimens that the held-out assignment of Figure~\ref{fig:aorta-pcal} leaves undecided.}
\label{fig:aorta}
\end{figure}

\experiment{Experiment 5: assignment to a cohort} We fit the populations of the two cohorts jointly, with shared hidden layers (Section~\ref{sec:cohort}). For illustration, Figure~\ref{fig:aorta} shows the populations of a fit to all twelve specimens with their labels; this fit enters neither experiment. Like the measured curves, the two populations differ in shape: the aneurysmal population is soft at small stretch and stiffens sharply, whereas the healthy one is stiffer at small stretch and stiffens gradually.

\begin{figure}[htbp]
\centering
\includegraphics[width=0.6\textwidth]{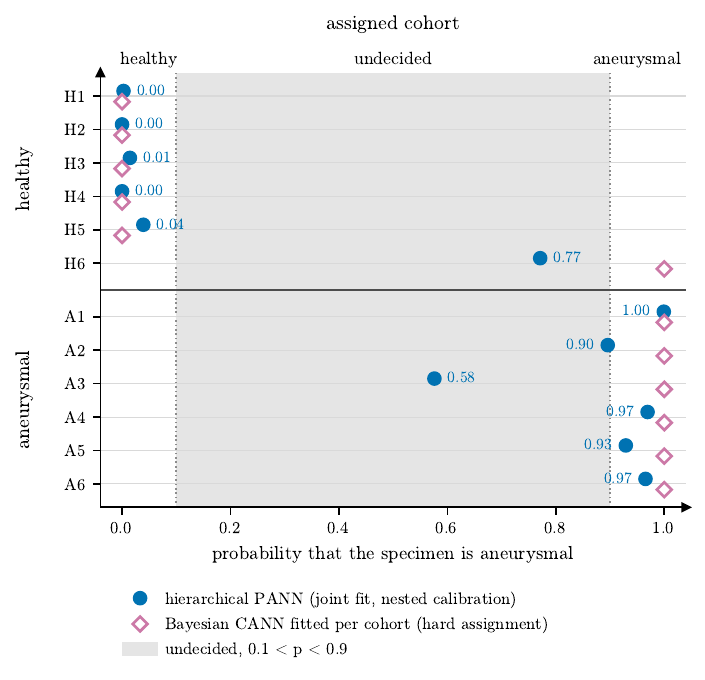}
\caption{Assignment of each held-out aorta specimen to a cohort (Experiment~5). Our probability that the specimen is aneurysmal comes from the log Bayes factor \eqref{eq:bayesfactor} of its circumferential curve under the joint fit of both cohorts without it, calibrated by nested leave-one-out (Section~\ref{sec:cohort}). The Bayesian CANN \cite{linka2025}, fitted per cohort, assigns the specimen to the cohort whose law gives its circumferential curve the larger predictive likelihood, which yields a hard assignment without a probability, drawn at $0$ or $1$.}
\label{fig:aorta-pcal}
\end{figure}

Each held-out specimen receives the calibrated probability of its cohort from the log Bayes factor \eqref{eq:bayesfactor} of its circumferential curve (Section~\ref{sec:cohort}). We call a specimen undecided if neither cohort reaches a probability of $0.9$. Figure~\ref{fig:aorta-pcal} shows the result. Ten of the twelve specimens are assigned to their true cohort with a probability of at least $0.9$ (A2 at the edge, $0.90$). The other two, H6 and A3, are undecided, and no specimen is assigned to the wrong cohort with confidence. The Bayesian CANN assigns $11$ of the $12$ specimens correctly, and assigns H6 to the aneurysmal cohort without a measure of how certain this is.

\begin{figure}[htbp]
\centering
\includegraphics[width=\textwidth]{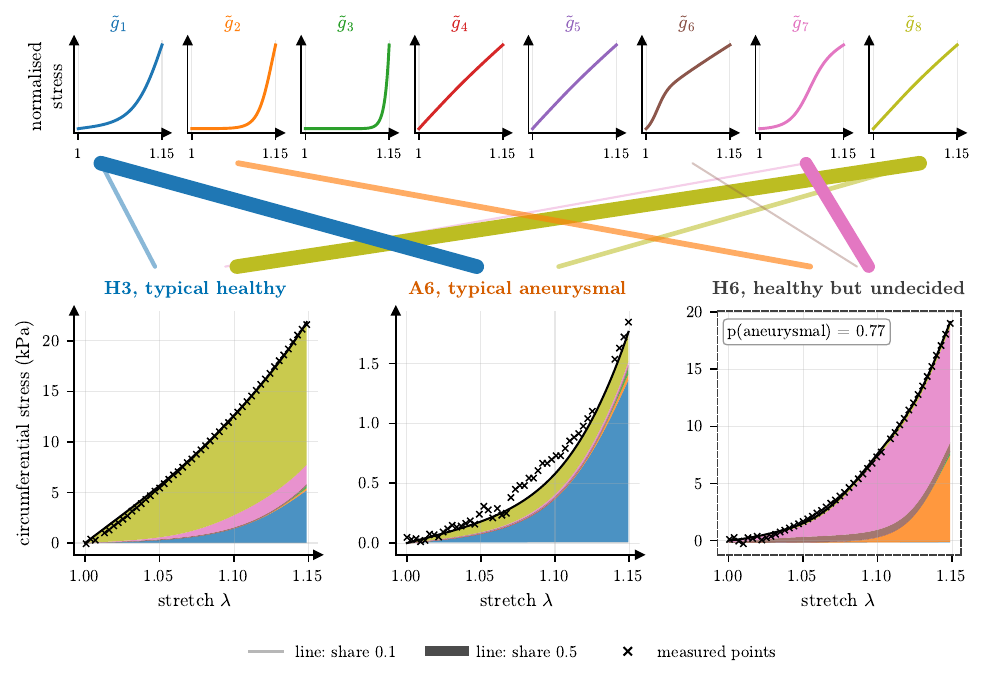}
\caption{Decomposition of the aorta curves into the shared features, for illustration (Experiment~5). The fit is that of Figure~\ref{fig:aorta}, to all twelve specimens with their labels. The top row shows the eight normalised features $\tilde g_j$ \eqref{eq:featnorm} shared by both cohorts, as circumferential stress per unit coefficient, ordered from the feature used most by the aneurysmal to that used most by the healthy specimens. The bottom row stacks the circumferential stress of three specimens by feature. H3 and A6 are the specimens whose shares are closest to the mean of their cohort, and H6 is the healthy specimen that the held-out assignment of Figure~\ref{fig:aorta-pcal} leaves undecided. The width of a line from a feature to a specimen is proportional to the posterior-mean share $\pi_j$ of the feature, and shares below $0.05$ are not drawn.}
\label{fig:aorta-features}
\end{figure}

To illustrate what distinguishes the cohorts, Figure~\ref{fig:aorta-features} decomposes the curves of the illustration fit of Figure~\ref{fig:aorta} into its features. The two cohorts share the features of the hidden layers and differ in how they weight them. The healthy specimens H1 to H5 draw most of their circumferential stress from $\tilde g_8$, which stiffens gradually, whereas most aneurysmal specimens draw theirs mainly from $\tilde g_1$ to $\tilde g_3$, which stiffen sharply at large stretch. H6 is the exception. Although its label is given and the prior on the spreads favours specimens that draw on the same features, the fit represents it mainly by $\tilde g_7$ and $\tilde g_2$, which H1 to H5 use little. H6 thus does not fit the healthy population even with its label, and lies between the cohorts, consistent with its undecided assignment above.

\experiment{Experiment 6: prediction from a partial test} Predicting with the model of Experiment~5 would require deciding on one cohort and discard the uncertainty of the assignment. We therefore predict the axial curve of each held-out specimen with one population, fitted to the eleven other specimens without their cohort labels (the unlabelled population) and conditioned on the circumferential curve (Section~\ref{sec:partial-test}). A single population suffices since specimens such as H6 suggest that the cohorts overlap rather than form two separate clusters. The Bayesian CANN, fitted per cohort with its own noise model, first assigns the specimen and predicts with the law of the assigned cohort. The no-population reference is a Bayesian PANN fitted to the circumferential curve alone, with its inputs normalised over the same range as those of the populations. We compare the $95\,\%$ posterior predictive bands, which include the noise,\footnote{Ours and the fit without a population use the noise level $\hat s$, the mean of the posterior median noise levels of the two cohort populations fitted without the specimen; the Bayesian CANN uses its own learned noise.} with the measured axial points and score them by relative error and coverage (\ref{app:evaluation}).

\begin{figure}[htbp]
\centering
\includegraphics[width=\textwidth]{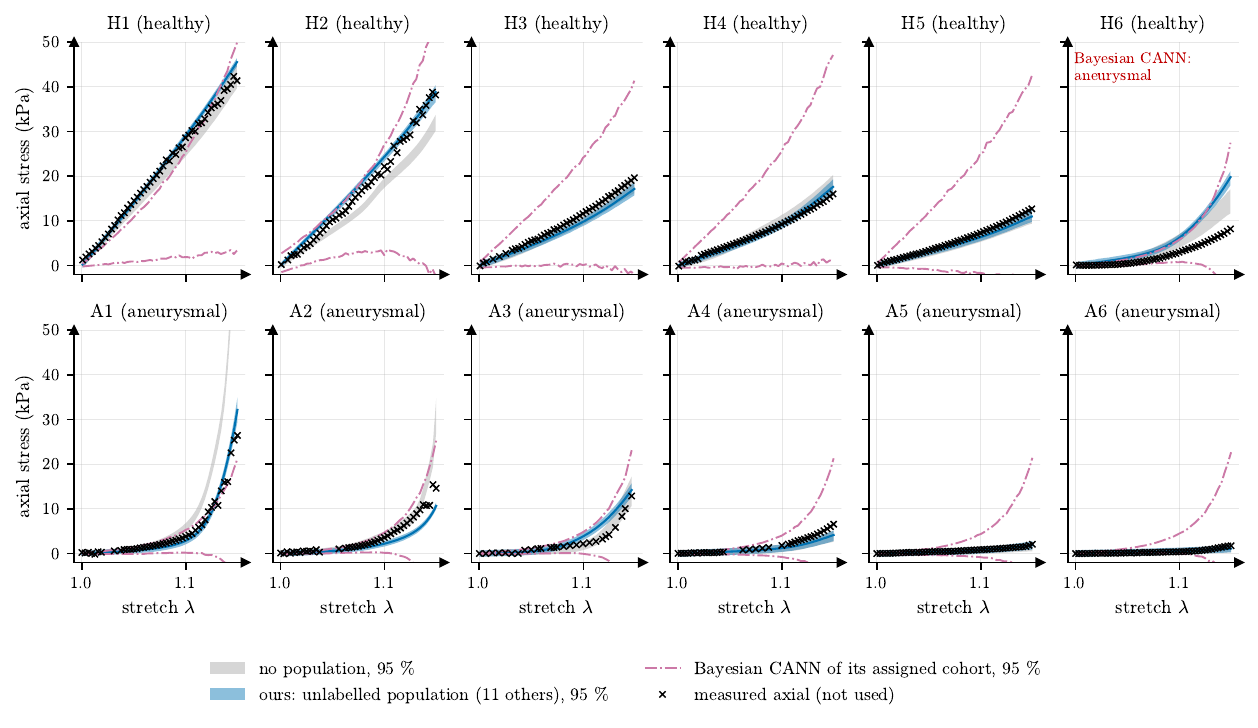}
\caption{Prediction of the untested axial curve of each held-out aorta specimen from its circumferential curve (Experiment~6). Our population is fitted to the eleven other specimens without their cohort labels and conditioned on the circumferential curve (Section~\ref{sec:partial-test}). The fit without a population is a Bayesian PANN (Section~\ref{sec:onespecimen}) fitted to the circumferential curve alone, and the Bayesian CANN \cite{linka2025} predicts with the law of the cohort that it assigns. All bands are $95\,\%$ posterior predictive bands and include the noise. No prediction uses the measured axial points, and the red label marks the specimen that the Bayesian CANN assigns to the wrong cohort. The figure shows the first of three restarts.}
\label{fig:predict-unl}
\end{figure}

Figure~\ref{fig:predict-unl} shows the predictions, and Table~\ref{tab:aorta-cp-summary} summarises them. Our unlabelled population predicts the axial curve most accurately, and its band follows the axial points of most specimens closely, healthy and aneurysmal alike. Its error is slightly lower than that of the fit without a population and about a third of that of the Bayesian CANN. Its band, however, covers only about $0.6$ of the axial points, below the nominal $0.95$, which is the failure of the complete test in Section~\ref{sec:newspecimen}. H6 is the clearest case, for which our band predicts an axial stiffening that the specimen does not show. The fit without a population covers more of the axial points, but is less accurate on average and fails badly on A1 and H6, since it has to extrapolate the axial direction from the circumferential curve alone.

The Bayesian CANN covers close to the nominal fraction of the axial points, and all of them for most specimens, but with a band about seven times as wide as ours. Its band has a similar width for every specimen of a cohort, whatever the specimen's stress level, and for most specimens it reaches from compressive stresses, which a tension test cannot produce, to several times the measured stress. This is the pooling trap of Section~\ref{sec:intro-pooling}: with one law per cohort, the fit attributes the differences between specimens to the noise, which then dominates the band.

\begin{table}[htbp]
\centering
\small
\begin{tabular}{@{}lccccc@{}}
\toprule
 & \multicolumn{2}{c}{relative error} & \multicolumn{2}{c}{coverage} & band width \\
 & mean & median & mean & median & (kPa) \\
\midrule
ours, unlabelled population & \textbf{0.12} & \textbf{0.08} & 0.61 & 0.62 & \textbf{1.6} \\
Bayesian CANN & 0.33 & 0.19 & \textbf{0.89} & \textbf{1.00} & 11.8 \\
no population & 0.15 & 0.10 & 0.74 & 0.85 & 2.2 \\
\bottomrule
\end{tabular}

\caption{Accuracy and calibration of the predicted axial curves of the held-out aorta specimens (Experiment~6, Figure~\ref{fig:predict-unl}). The relative error (\ref{app:evaluation}) and the coverage of the measured axial points by the $95\,\%$ predictive band are given as mean and median over the twelve specimens and three restarts, and the band width as a mean. Bold marks the lowest error, the coverage closest to the nominal $0.95$ and the narrowest band.}
\label{tab:aorta-cp-summary}
\end{table}

\FloatBarrier

\clearpage
\section{Discussion and conclusions}
\label{sec:discussion}

We proposed a Bayesian PANN for one specimen and its hierarchical extension to a population of laws. For one specimen, the priors keep the learned law smooth and plausible from $2\,\%$ noise on, where the deterministic PANN follows the noise, and its restarts nearly coincide. On Treloar's rubber, it reaches about the accuracy that the PANN architecture allows, and its predictive band covers every held-out point.

The main use of the hierarchical model is the prediction of a new specimen from a partial test (\ref{req:pred}). After a short uniaxial test, the predicted band covers about $90\,\%$ of the true curves of the untested load cases, close to a reference that uses the correct form of the law, and with a dozen specimens in the population, it approaches this reference. The competing methods either miss the untested response or bound it only with much wider bands.

The population also meets the other requirements of Section~\ref{sec:intro-req}. With six specimens, it is already close to the true population, whereas the competitors remain farther from it up to twelve specimens, and the closest of them, the Gaussian CANN, catches up only at $24$ (\ref{req:data}). At a higher noise level, the population is inferred as before, whereas methods without a separate noise model read the noise as variation between specimens (\ref{req:noise}). On the aortas, the cohort populations assign no held-out specimen to the wrong cohort with confidence and leave the specimens between the cohorts undecided, whereas the pooled laws assign a specimen to the wrong cohort without a measure of certainty (\ref{req:cohort}). One population fitted without the labels predicts the untested axial curve more accurately than a fit to the specimen alone or the pooled law of the assigned cohort (Table~\ref{tab:aorta-cp-summary}), but its band is too narrow.

The main limitation is overconfidence for very few specimens. With three specimens, the smallest number from which the spread of a population can be estimated at all, the learned population should be wider than the true one, but ours stays at about the true width. Similarly, a long and precise test of a new specimen locates it more sharply within the population than is justified, so that the band of the untested load cases misses the true curve more often than a $95\,\%$ band should. The too narrow bands on the aortas are the same effect. Future work may address this in the model, with a population that is not Gaussian; in the prior, with priors on the spreads that favour wider populations when the specimens are few; or in the inference, with an approximation of the posterior that is more flexible than the Gaussian $q$,
e.g.\ a mixture that captures several modes \cite{franck_multimodal_2017}, which is affordable since the population lives in few coordinates.
Instead of resampling, the population could also be refitted with the new specimen included, which targets the same posterior without being limited to the draws of the fitted population.

A second limitation concerns the prior. We used one general prior unchanged in all experiments. It supplies what few or noisy data leave open and keeps the learned laws plausible, but it is not the best choice for every material, and where prior knowledge about the material is available, the prior should be chosen accordingly.

Since every draw of the population is an admissible PANN (\ref{req:adm}), propagating the population through a finite element simulation would turn the stress bands of this paper into bounds on a structural quantity of interest, such as the wall stress of an aneurysm or the deflection of a component, and a simulation surrogate would keep the cost affordable. For recyclates, one population per material stream would assign a new batch to its stream and predict its untested response \cite{scholz2026thermoplastic}. In both settings, a population that carries its own uncertainty turns the few specimens a study can afford into predictions about the specimens it cannot test.

\clearpage
\appendix
\makeatletter
\@addtoreset{figure}{section}
\@addtoreset{table}{section}
\@addtoreset{equation}{section}
\renewcommand{\p@section}{Appendix~}
\renewcommand{\p@subsection}{Appendix~}
\makeatother
\renewcommand{\thefigure}{\thesection.\arabic{figure}}
\renewcommand{\thetable}{\thesection.\arabic{table}}
\renewcommand{\theequation}{\thesection.\arabic{equation}}
\section{Details of the PANN}
\label{app:pann}

\paragraph{Correction terms} The correction terms of Section~\ref{sec:pann} are those of \cite{linden2023neural},
\begin{subequations}
\label{eq:corrections}
\begin{align}
  \Psi_{\mathrm{corr}} &= \Psi_{\mathrm{energy}} + \Psi_{\mathrm{stress}} + \Psi_{\mathrm{growth}}, \label{eq:psi-total} \\
  \Psi_{\mathrm{energy}} &= -\Psi_{\mathrm{NN}}\big|_{\bs{C} = \bs{I}}, \label{eq:psi-energy} \\
  \Psi_{\mathrm{stress}} &= -\mathfrak{n}\,(J - 1), \qquad
  \mathfrak{n} = 2\Big(\frac{\partial \Psi_{\mathrm{NN}}}{\partial I_1} + 2\,\frac{\partial \Psi_{\mathrm{NN}}}{\partial I_2} + \frac{\partial \Psi_{\mathrm{NN}}}{\partial I_3} - \frac{\partial \Psi_{\mathrm{NN}}}{\partial I_1^*}\Big)\Big|_{\bs{C} = \bs{I}}, \label{eq:psi-stress} \\
  \Psi_{\mathrm{growth}} &= \Big(J + \frac{1}{J} - 2\Big)^2. \label{eq:psi-growth}
\end{align}
\end{subequations}
The first term makes the energy and the second the stress vanish in the undeformed state, and the third lets the energy grow without bound for $J \to 0$ and $J \to \infty$.

\paragraph{Standardisation of the invariants} The values $\mathcal{I}^0_m$ and $\Delta\mathcal{I}_m$ in \eqref{eq:standardise} are computed once, stored with the trained model and applied unchanged to every input at test time. Since the prior of Section~\ref{sec:onespecimen} relies on inputs that are centred and of order one, the standardisation is part of the prior. Instead of the training states, $\mathcal{I}^0_m$ and $\Delta\mathcal{I}_m$ can also be taken from the range of states expected at test time, or from a conservative standard choice such as the undeformed state, $\mathcal{I}^0 = (3, 3, 1, -2)$, with $\Delta\mathcal{I}_m$ the largest deviation from it over UT, ET and PS (Table~\ref{tab:loadcases}) up to a maximum stretch of, e.g., $5$ for elastomers or $1.3$ for soft tissue.

\section{Evaluation scores}
\label{app:evaluation}

All bands that enter a score are central $95\,\%$ intervals $[\ell, u]$, bounded at every stretch by the $2.5\,\%$ and $97.5\,\%$ quantiles of the draws, for the single specimen of Section~\ref{sec:onespecimen-results} as for the populations. For a population, stresses are scored in the scaled units of Section~\ref{sec:setup-data}, so that the scores are comparable across load cases and methods. For a single specimen, errors and widths are divided by the peak stress of the true curve of the load case.

\paragraph{Error} The relative error of a predicted curve is the root mean square distance of the posterior median from the true curve (synthetic data) or from the measured points (real data), divided by the peak stress of that curve. For the single specimen of Figures~\ref{fig:single} and~\ref{fig:restarts}, the distance is evaluated at $200$ equally spaced stretches over the range of the held-out PS test, and the median is taken over $2000$ posterior draws; for the deterministic PANN its fitted curve takes the place of the median.

\paragraph{Coverage and width} The coverage of a curve is the fraction of its stretch points that lie inside the band. For a population, it is this fraction over the curves of $200$ new specimens drawn from the true population. The band width is $u - \ell$, averaged over the stretch points. For a population, it is reported relative to the width of the $95\,\%$ band of the true population, estimated from $2\cdot 10^{4}$ specimens, as the ratio of the two averaged widths per load case.

\paragraph{Energy distance} The distance between the new specimens of a fitted population, $X$, and those of the true population, $Y$, is the energy distance \cite{szekely2013energy}
\begin{equation}
  D(X, Y) = 2\,\E\|X - Y\| - \E\|X - X'\| - \E\|Y - Y'\|,
  \label{eq:edist}
\end{equation}
where $X'$ and $Y'$ are independent copies of $X$ and $Y$. Every specimen is represented by its scaled stresses at the $30$ stretch points of each of the three load cases, and $\|\cdot\|$ is the root mean square over these $90$ values. The expectations are means over all pairs of $2000$ new specimens of the fit and $2000$ specimens of the true population, and we report $D$ itself, not $\sqrt{D}$. It is zero if and only if the two populations coincide.

\paragraph{Energy screen} Polyconvexity holds for every draw by construction, a non-negative energy does not. For the single-specimen fits of Section~\ref{sec:onespecimen-results}, whose law is compressible, we evaluate the energy of $2000$ posterior draws per fit along the volumetric path $\bs{C} = \lambda^{2}\bs{I}$ at $2000$ logarithmically spaced stretches $\lambda \in [0.05, 20]$, which do not include $\lambda = 1$. The fits are the five data realisations and the $20$ restarts of Figure~\ref{fig:restarts} per noise level, and a draw counts as failed if $\Psi < 0$ at any of these stretches, without a tolerance.

\section{Setup of the experiments}
\label{app:setup}

\paragraph{Load cases and noise} Unlike the tests of \cite{linden2023neural}, the lateral stretches of the compressible laws are prescribed rather than solved for, so that every posterior draw is evaluated at the same states. UT and ET are therefore a uniaxial and an equibiaxial stretch rather than tension tests, and we keep the abbreviations for both. Within a data set, the standard-normal draws behind $\epsilon_i$ in \eqref{eq:noise-synth} are the same at every noise level, scaled by $\rho$.

\paragraph{Settings and computation times} Table~\ref{tab:setup} lists the settings that the formulation leaves open, for each model. The restarts of the aorta fits (Section~\ref{sec:aorta}) differ in the random seed of the starting fits (Section~\ref{sec:inference}) and of the Monte Carlo draws with which the gradient of the ELBO is estimated. All computations ran on one workstation with an AMD Ryzen Threadripper PRO 5995WX processor and NVIDIA GeForce RTX 4090 graphics cards, and Table~\ref{tab:times} lists the computation times.

\begin{table}[htbp]
\centering
\small
\begin{tabular}{@{}p{0.42\textwidth}p{0.5\textwidth}@{}}
\toprule
Hyperparameter & Value \\
\midrule
\multicolumn{2}{@{}l}{\emph{Network and priors (both models)}} \\
Hidden layers $\times$ neurons & $2 \times 8$ \\
Activation & softplus \\
Inputs & standardised invariants \eqref{eq:standardise} \\
Prior of the hidden layers & fan-in widened by $\kappa = 3$ \eqref{eq:single} \\
Prior of the (mean) output layer & $\mathrm{Dirichlet}(\tfrac12\bs{1}_N)$; $a \sim \mathcal{N}(\bar a, s_a^2)$ \eqref{eq:single-out} \\
\multicolumn{2}{@{}l}{\emph{Bayesian PANN (one specimen)}} \\
Stress bounds for $\bar a$, $s_a$ & S-shape law: $10^{-3}$ to $10^{3}$; Treloar: $1$\,Pa to $1$\,GPa \\
Unknowns & $D = 120$ \\
Approximate posterior & Gaussian, full covariance \\
Initialisation & best of $5$ MAP fits; $\bs{L} = 0.05\,\bs{I}$ \\
Optimiser & Adam, gradient norm clipped at $10$ \\
Learning rate & $10^{-2} \to 10^{-3}$, exponential decay \\
Steps & $4 \cdot 10^{4}$ \\
Draws per step $N_{\mathrm{mc}}$ & $32$ \\
Draws for the bands $N_{\mathrm{pred}}$ & $2000$ \\
\multicolumn{2}{@{}l}{\emph{Hierarchical Bayesian PANN}} \\
Stress bounds for $\bar a$, $s_a$ & $10^{-3}$ to $10^{3}$: $\bar a = 0$, $s_a = 3.99$ \\
Number of modes $r$ & $5$ \\
Priors of the scales & $P(\gamma_l > 1) = P(\tau_a > 0.3) = P(\tau_z > 0.3) = 0.05$ \eqref{eq:priors} \\
Unknowns & $127$ shared $+\ 8K$ specimen coordinates \\
Modes $\bs{u}_1, \dots, \bs{u}_r$ & $5$ unit vectors in $\mathbb{R}^{8}$, point estimates \\
Approximate posterior, shared & rank $10$ + diagonal \\
Approximate posterior, specimens & full covariance, $8 \times 8$ each \\
Initialisation & structured start, best of $8$ MAP fits \\
Optimiser, learning rate & as for the Bayesian PANN \\
Steps & $10^{5}$ \\
Draws per step $N_{\mathrm{mc}}$ & $128$ \\
Draws for the bands & $1000$ new specimens \\
\multicolumn{2}{@{}l}{\emph{Deterministic PANN and MAP fits}} \\
Optimiser & Adam \\
Learning rate & $10^{-2} \to 10^{-4}$, exponential decay \\
Steps & $10^{4}$ \\
\bottomrule
\end{tabular}
\caption{Settings of the fits, which apply unless an experiment states otherwise. The number of unknowns $D$ of the Bayesian PANN is that of the compressible network with four invariants, $D = 4\cdot 8 + 8 + 8\cdot 8 + 8 + 8$.}
\label{tab:setup}
\end{table}

\begin{table}[htbp]
\centering
\small
\begin{tabular}{@{}p{0.58\textwidth}ll@{}}
\toprule
Model or procedure & Hardware & Time \\
\midrule
Bayesian PANN, one fit (incl.\ $5$ MAP starts) & GPU & $9$\,min \\
Hierarchical Bayesian PANN, one fit ($K = 6$, $10^{5}$ steps) & GPU & $25$\,min \\
Prediction from a partial test (SIR, $2\cdot 10^{6}$ proposals), $75$ tests of one population & GPU & $33$\,s \\
Cohort assignment, leave-one-out (aortas, $12$ specimens, both cohorts) & GPU & $25$\,min \\
\midrule
Deterministic PANN, $5$ restarts & GPU & $1$\,min \\
Known-law ceiling (NUTS, $4$ chains) & CPU, $4$ cores & $9$\,s \\
Ta\c{c} et al.\ \cite{tac2024generative}, one network width ($10$ in their protocol) & CPU, $2$ threads & $7$\,min \\
Gaussian CANN \cite{mcculloch2025discovering} & CPU, $2$ threads & $14$\,min \\
Bayesian CANN \cite{linka2025}, synthetic data / aortas & CPU, $2$ / $4$ threads & $5$ / $8$\,min \\
\bottomrule
\end{tabular}
\caption{Computation times of the fits and procedures, as wall-clock times measured alone on the workstation described above, where GPU denotes one RTX 4090. Our models are listed above the line, and the references and competitors below it. The time of the known-law ceiling is per fit, from the runs reported here with $16$ fits in parallel.}
\label{tab:times}
\end{table}

\FloatBarrier

\section{Competitors and the known-law ceiling}
\label{app:competitors}

We use the published code of each method: that of Ta\c{c} et al.\ \cite{tac2024generative} from \url{https://github.com/tajtac/node_diffusion} at commit \texttt{b8dbbb0}, and that of the Gaussian CANN \cite{mcculloch2025discovering} and the Bayesian CANN \cite{linka2025} from \url{https://github.com/LivingMatterLab/CANN} at commit \texttt{a639ea9}. Our adapted versions will be included in our repository. Each code is written for its authors' data. The changes below are those without which it would not run on our data or could not be compared, and none of them changes a method.
\begin{itemize}
  \item All methods receive the stresses divided by one global constant, so that they are of order one as in the authors' examples, which their loss weights and priors assume. Their draws are scaled back for scoring.
  \item All methods run for at least as many optimiser steps as in the authors' examples. Our data sets are smaller, so the same number of epochs would mean far fewer steps.
  \item Bayesian CANN: the invariants enter as $(I - 3)/\max_{\mathrm{train}}(I - 3)$, as in the code of the Gaussian CANN \cite{mcculloch2025discovering}, since its exponential terms overflow at our stretches otherwise. It is trained on all three load cases, and each curve of its population uses one draw of all weights.
  \item Gaussian CANN: our examples are isotropic, so we use only the terms in $I_1$ and $I_2$ of its library.
  \item Ta\c{c} et al.: their loss fits both in-plane Cauchy stresses, $\sigma_{11}$ and $\sigma_{22}$. In UT and ET, $\sigma_{22}$ follows from the data, as $0$ and as the measured $\sigma_{11}$. In PS, along whose second in-plane direction the stretch is held at one, our data do not contain $\sigma_{22}$, so we compute it from the true law and add noise of the same level. This method therefore receives one observable more than ours.
\end{itemize}

\paragraph{Prediction for a new specimen}
Of the three methods, only Ta\c{c} et al.\ define a prediction for a new specimen from its measurements, and we use it as published \cite{tac2024generative}, with the conditioning width $\varsigma = 0.04$ that their code sets, and with $\varsigma = 0.5$ on two fits as a sensitivity check. This more than tenfold looser width widens the band only mildly: for tests up to $\lambda_{\max} = 2$ and $3$ it then covers $0.16$ to $0.31$ of the true curves of ET and PS, against fewer than $0.2$ with the published width, so we keep the published value. The width is fixed and not computed from the data. The one setting that their code selects from the data, the number of specimen-specific parameters, is selected by the energy distance to a statistic of the training specimens only. The Gaussian CANN and the Bayesian CANN fit a population but provide no update of it by the measurements of one specimen. For both we add this update, in the same way as for our model: the population fitted by the method is the prior of the new specimen, and the likelihood of its partial test is the Gaussian likelihood \eqref{eq:sir} with the known noise level $\sigma_\epsilon$. These updates are our extensions of the two methods.

\emph{Bayesian CANN.} Its population is the variational posterior $q(\bs{w})$ of its weights. We condition it exactly as our population, by sampling-importance-resampling (Algorithm~\ref{alg:sir}) with $10^{5}$ draws from $q(\bs{w})$ in place of the draws of our population, so that both are conditioned by the same procedure.

\emph{Gaussian CANN.} Its population is the generating model of \cite{mcculloch2025discovering}, in which a specimen scales every term of the fitted law by its own random factor,
\begin{equation}
  P_{11}(\lambda) = \sum_{i=1}^{n} t_i(\lambda)\,\big(1 + (\bs{K}\bs{z})_i\big), \qquad \bs{z} \sim \mathcal{N}(\bs{0}, \bs{I}_n),
  \label{eq:gcann-gen}
\end{equation}
where $t_i(\lambda)$ is the stress of term $i$ at the fitted weights, $\bs{K}$ the fitted lower-triangular covariance factor and $n = 8$ the number of terms; one $\bs{z}$ is drawn per specimen and shared by its load cases. This is the prior of the new specimen. Since the stress is linear in $\bs{z}$, $P_{11}(\lambda) = m(\lambda) + \bs{a}(\lambda)^{\top}\bs{z}$ with $m = \sum_i t_i$ and $\bs{a} = \bs{K}^{\top}\bs{t}$, the posterior given the partial test $\bs{y} = (y_1, \dots, y_M)$ at the stretches $\lambda_1, \dots, \lambda_M$ is Gaussian \cite{bishop2024deep},
\begin{equation}
  \begin{gathered}
  \bs{z} \mid \bs{y} \sim \mathcal{N}(\bs{\mu}_z, \bs{\Sigma}_z), \\
  \bs{\Sigma}_z = \big(\bs{I}_n + \sigma_\epsilon^{-2}\bs{A}^{\top}\bs{A}\big)^{-1}, \qquad
  \bs{\mu}_z = \sigma_\epsilon^{-2}\,\bs{\Sigma}_z \bs{A}^{\top}(\bs{y} - \bs{m}),
  \end{gathered}
  \label{eq:gcann-post}
\end{equation}
where the rows of $\bs{A}$ are $\bs{a}(\lambda_j)^{\top}$ and $\bs{m} = (m(\lambda_j))_j$. A draw with $1 + (\bs{K}\bs{z})_i < 0$ gives term $i$ a negative weight, i.e.\ a law that is not admissible. We draw $2\cdot 10^{4}$ samples from \eqref{eq:gcann-post}, discard those with a negative factor on a term with non-zero stress, and take the band from the quantiles of the remaining ones; the figures of Section~\ref{sec:newspecimen} and \ref{app:newspec} show this variant. We discard them so that the Gaussian CANN, like our model, predicts only admissible laws; McCulloch et al.\ keep such draws \cite{mcculloch2025discovering}. Keeping them would not change our results significantly. Discarding requires knowing which terms carry stress, which is not possible in every fit, and it leaves a specimen without a band if too few draws remain.

\emph{Bayesian CANN for the aortas.} The published code of the Bayesian CANN covers isotropic materials only. For the aortas, we implement the transversely isotropic model of \cite{linka2025} in this code to the best of our knowledge, with the terms in $I_4$ and $I_5$ and the stress of the biaxial tests with two fibre families as described there, and keep everything else as published. Two settings differ. Training runs for $4\cdot 10^{4}$ optimiser steps, as on the synthetic data, i.e.\ $4000$ epochs in place of the $2500$ of their code, and the two directions of a cohort, which have different numbers of points, are padded to equal length with samples of zero weight, since their code requires equal sample counts.

\paragraph{Known-law ceiling} The ceiling is the hierarchical model
\begin{align}
  (\log c_{1,k}, \log a_k) &\sim \mathcal{N}\big(\bs{\mu}, \mathrm{diag}(\bs{d})\,\bs{R}\,\mathrm{diag}(\bs{d})\big), \nonumber \\
  \bs{\mu} &\sim \mathcal{N}(\bs{m}, \bs{I}), \qquad d_j \sim \mathcal{N}^{+}(0, 1), \qquad \bs{R} \sim \mathrm{LKJ}(\eta = 1),
\end{align}
with the law \eqref{eq:upturninc}, the likelihood of the same data with the known noise level, and sampling by NUTS \cite{hoffman2014nuts}. Here $\mathrm{LKJ}(\eta = 1)$ is the uniform distribution over $2 \times 2$ correlation matrices \cite{lewandowski2009generating}. The priors are weakly informative: a normal prior with unit standard deviation on each component of the mean, half-normal priors with unit scale on the two standard deviations, and a uniform prior on the correlation. The prior of the mean is centred at the true mean $\bs{m}$ of \eqref{eq:upturninc-pop}.

\paragraph{Behaviour of the competitors} We give our interpretation of three observations of Section~\ref{sec:competitors}.

\emph{Funnel of the Gaussian CANN.} A new specimen of the Gaussian CANN scales every term of the law by its own random factor $1 + (\bs{K}\bs{z})_i$ \eqref{eq:gcann-gen}. Its library has no $(I_1 - 3)^3$ term, so the upturn is carried by terms whose stress grows quickly with the stretch, and the same relative scatter of these terms then gives a band that widens quickly at large stretches. A factor can also be negative, and the stress of such a draw falls at large stretches, which pulls the lower edge of the band down.

\emph{Wrong stiffness--upturn correlation.} The positive correlation is a property of the specimens, which a method can learn only from which measured points belong to the same specimen. If $\log c_1$ and $\log a$ in \eqref{eq:upturninc-pop} were independent, the correlation \eqref{eq:corr-su} would be $-0.43$, since a larger linear term raises the stiffness and dilutes the relative upturn; the true $+0.46$ requires the strong positive correlation $0.8$ between the two terms. The Gaussian CANN fits the mean and the variance of the stress at each stretch, treating every measured point as an independent draw \cite{mcculloch2025discovering}, so it never sees which points belong to the same specimen. The correlation between its terms is constrained only weakly, through the variance at each stretch, and in the fits of this study the two active terms, a nearly linear and a stiffening one, have a correlation between $-0.56$ and $-0.34$.

\emph{Higher noise.} The competitors have no separate place for the measurement noise, so that it enters the learned scatter between specimens.

\section{Populations at higher noise}
\label{app:pop-noise}

\begin{figure}[htbp]
\centering
\includegraphics[width=\textwidth]{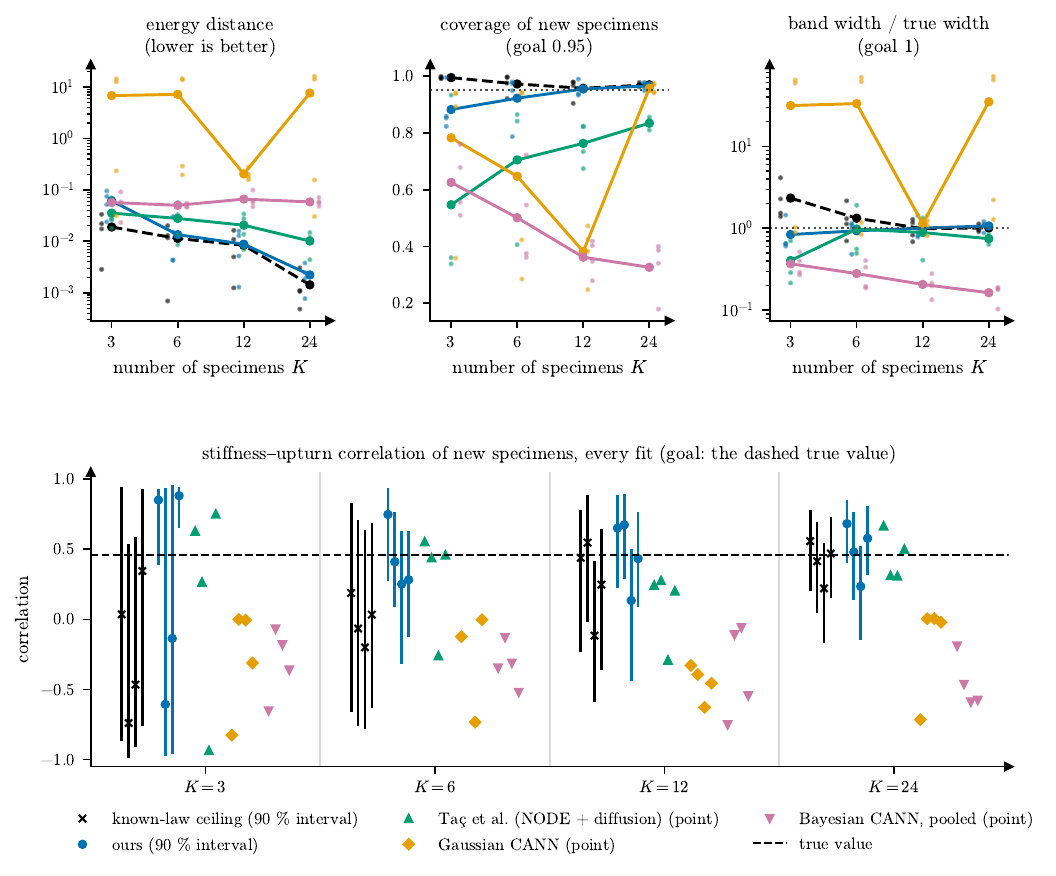}
\caption{Dependence of the learned populations on the number of specimens $K$ at $5\,\%$ noise (\ref{app:pop-noise}), shown as in Figure~\ref{fig:comp-metrics}.}
\label{fig:comp-metrics-5}
\end{figure}

Figure~\ref{fig:comp-metrics-5} repeats Figure~\ref{fig:comp-metrics} at $5\,\%$ noise.

\FloatBarrier

\section{New specimen from a population of twelve specimens}
\label{app:newspec}

\begin{figure}[htbp]
\centering
\includegraphics[width=\textwidth]{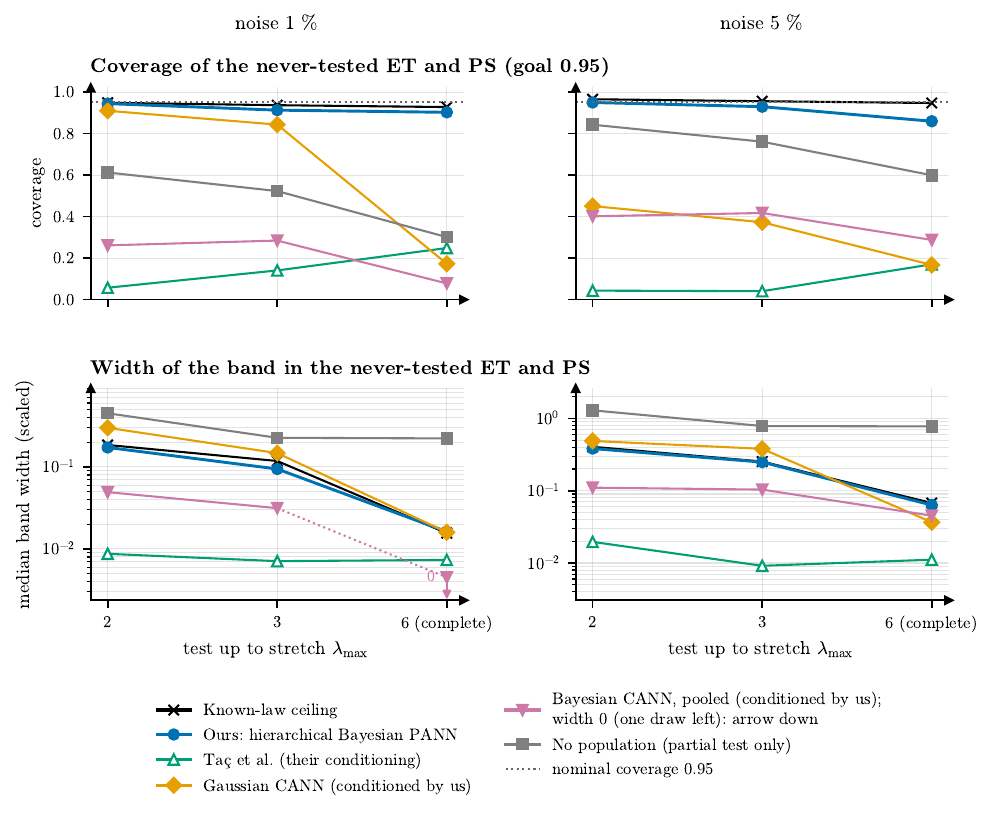}
\caption{Prediction of new specimens from partial uniaxial tests for populations from $K = 12$ specimens (\ref{app:newspec}), shown as in Figure~\ref{fig:ns-summary}.}
\label{fig:ns-summary-k12}
\end{figure}

Figure~\ref{fig:ns-summary-k12} repeats Figure~\ref{fig:ns-summary} for the populations learned from $K = 12$ specimens.

\FloatBarrier

\section{Aorta data}
\label{app:aortadata}

\begin{figure}[htbp]
\centering
\includegraphics[width=\textwidth]{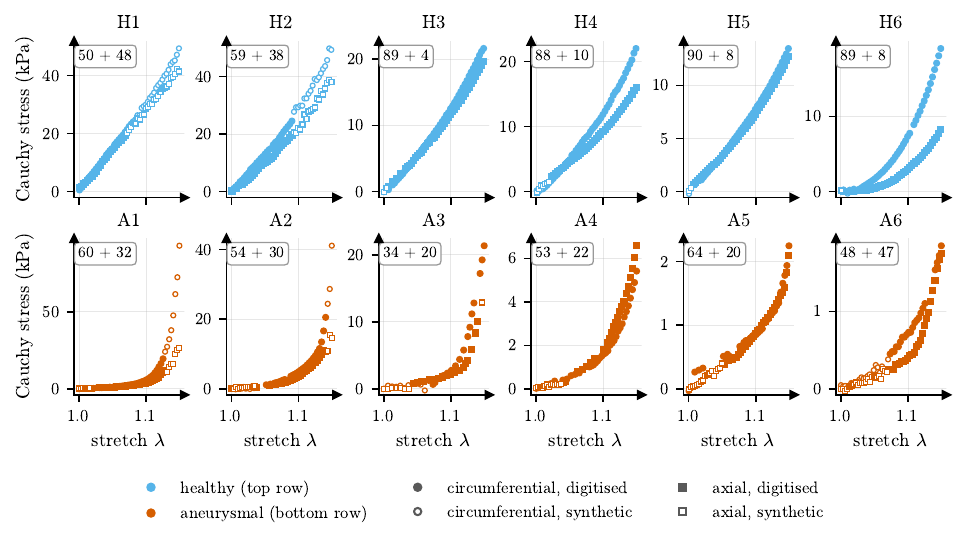}
\caption{Aorta data used in Experiments~5 and~6, from equibiaxial tests of six healthy and six aneurysmal specimens \cite{niestrawska2016microstructure} in the form reported by Linka et al.\ \cite{linka2025}. Each specimen gives a circumferential and an axial curve. The points were digitised from published figures, and curves that end before the stiff regime are completed up to a stretch of $1.15$ with synthetic points from a per-specimen fit of a Holzapfel--Demiray law. The boxes give the number of digitised and synthetic points.}
\label{fig:aortadata}
\end{figure}

We use the data of Niestrawska et al.\ \cite{niestrawska2016microstructure} in the form reported by Linka et al.\ \cite{linka2025}. The tissue data are not openly available, so we digitised $778$ points from published figures. The digitised curves of several specimens end at small stretches and do not reach the stiff regime, so we added $287$ synthetic points beyond or before the digitised range, up to a stretch of $1.15$, from per-specimen fits of a Holzapfel--Demiray law (specimen H6, which shows a toe region, from the dispersed fibre law of \cite{niestrawska2016microstructure}). The synthetic points carry Gaussian noise of about the scatter of the specimen's digitised points around its fitted law. Figure~\ref{fig:aortadata} shows the data and marks digitised and synthetic points.

\FloatBarrier

\section{Transversely isotropic PANN for the aortas}
\label{app:ti}

For the aortas we use the transversely isotropic, incompressible PANN of Linden et al.\ \cite{linden2023neural} with one fibre direction $\bs{a}_0 = \bs{e}_1$ along the circumferential axis and the structural tensor $\bs{G} = \bs{a}_0 \otimes \bs{a}_0$. Besides the isotropic invariants $I_1$ and $I_2$, the network receives
\begin{equation}
  I_4 = \mathrm{tr}(\bs{C}\bs{G}), \qquad I_5 = \mathrm{tr}(\mathrm{cof}(\bs{C})\,\bs{G}),
  \label{eq:ti-invariants}
\end{equation}
while $I_3 = 1$ and $I_1^*$ are constant and dropped. The ICNN \eqref{eq:icnn}, its non-negative weights and the corrections \eqref{eq:corrections} are those of Section~\ref{sec:pann} and \ref{app:pann}, with the additional term $p\,(I_4 - 1) + q\,(I_5 - 1)$, $p = \max(-x, 0)$, $q = \max(x, 0)$ and $x = (\partial\Psi_{\mathrm{NN}}/\partial I_4 - \partial\Psi_{\mathrm{NN}}/\partial I_5)|_{\bs{C} = \bs{I}}$, which makes the stress vanish in the undeformed state \cite{linden2023neural}. The tests are equibiaxial, $\bs{F} = \mathrm{diag}(\lambda, \lambda, \lambda^{-2})$, and the pressure of the incompressibility constraint follows from $\sigma_{33} = 0$, so that
\begin{equation}
  \sigma_{ii} = s_i - s_3, \qquad s_i = 2\big[\psi_1 \lambda_i^2 + \psi_2 (I_1 \lambda_i^2 - \lambda_i^4) + \psi_4\, G_{ii} \lambda_i^2 + \psi_5\, I_5 (1 - G_{ii})\big],
  \label{eq:ti-stress}
\end{equation}
where $\psi_j = \partial\Psi/\partial I_j$, $\lambda_1 = \lambda_2 = \lambda$, $\lambda_3 = \lambda^{-2}$, and $i = 1$ is the circumferential and $i = 2$ the axial direction; this is the stress of Linka et al.\ \cite{linka2025} with a single fibre family. The population, the priors and the inference are those of Section~\ref{sec:method}.

\section*{Funding}
The support of the Deutsche Forschungsgemeinschaft (DFG) through Project Number 499746055 is gratefully acknowledged.

\section*{Author contributions}
\textbf{Vincent Scholz:} Conceptualization, Methodology, Software, Validation, Formal analysis, Investigation, Data curation, Visualization, Project administration, Writing -- original draft, Writing -- review \& editing.\\
\textbf{Phaedon-Stelios Koutsourelakis:} Funding acquisition, Resources, Supervision, Writing -- review \& editing.

\section*{Competing interests}
The authors declare that they have no known competing financial interests or personal relationships that could have appeared to influence the work reported in this paper.

\section*{Data availability}
The code of the proposed method, the adapted versions of the published methods used for comparison and the synthetic data sets will be made available upon publication at \url{https://github.com/pkmtum/Hierarchical-Bayesian-PANN}. The data of Treloar's rubber are those of \cite{treloar1944stress}. The aorta data are those of Niestrawska et al.\ \cite{niestrawska2016microstructure} in the form reported by Linka et al.\ \cite{linka2025}, digitised from the published figures; we do not have permission to redistribute them, and requests for these data should be directed to the original authors.

\section*{Use of generative AI}
During the preparation of this work the authors used Claude (Anthropic, model Claude Opus 5.5) and the Claude Code harness (with the same model) to assist in writing the code and to polish the formulation of the text. After using these tools, the authors reviewed and edited the content as needed and take full responsibility for the content of the publication.

\bibliographystyle{unsrtnat}
\bibliography{main}

@article{scholz_illusion_2026,
        title = {The illusion of fit: {Spatially} resolved assessment of constitutive model validity in elastography and physics-based inverse problems},
        volume = {461},
        issn = {0045-7825},
        shorttitle = {The illusion of fit},
        url = {https://www.sciencedirect.com/science/article/pii/S0045782526004810},
        doi = {10.1016/j.cma.2026.119208},
        urldate = {2026-11-06},
        journal = {Computer Methods in Applied Mechanics and Engineering},
        author = {Scholz, Vincent C. and Koutsourelakis, P. S.},
        month = nov,
        year = {2026},
        pages = {119208},
}

@article{franck_multimodal_2017,
        title = {Multimodal, high-dimensional, model-based, {Bayesian} inverse problems with applications in biomechanics},
        volume = {329},
        issn = {0021-9991},
        url = {http://www.sciencedirect.com/science/article/pii/S002199911630537X},
        doi = {10.1016/j.jcp.2016.10.039},
        urldate = {2017-03-07},
        journal = {Journal of Computational Physics},
        author = {Franck, I. M. and Koutsourelakis, P. S.},
        month = jan,
        year = {2017},
        pages = {91--125},
}

@article{linden2023neural,
  author  = {Linden, Lennart and Klein, Dominik K. and Kalina, Karl A. and Brummund, J{\"o}rg and Weeger, Oliver and K{\"a}stner, Markus},
  title   = {Neural networks meet hyperelasticity: A guide to enforcing physics},
  journal = {Journal of the Mechanics and Physics of Solids},
  volume  = {179},
  pages   = {105363},
  year    = {2023},
  doi     = {10.1016/j.jmps.2023.105363}}

@article{klein2022polyconvex,
  author  = {Klein, Dominik K. and Fern{\'a}ndez, Mauricio and Martin, Robert J. and Neff, Patrizio and Weeger, Oliver},
  title   = {Polyconvex anisotropic hyperelasticity with neural networks},
  journal = {Journal of the Mechanics and Physics of Solids},
  volume  = {159},
  pages   = {104703},
  year    = {2022},
  doi     = {10.1016/j.jmps.2021.104703}}

@inproceedings{amos2017input,
  author    = {Amos, Brandon and Xu, Lei and Kolter, J. Zico},
  title     = {Input Convex Neural Networks},
  booktitle = {Proceedings of the 34th International Conference on Machine Learning},
  editor    = {Precup, Doina and Teh, Yee Whye},
  series    = {Proceedings of Machine Learning Research},
  volume    = {70},
  pages     = {146--155},
  publisher = {PMLR},
  year      = {2017},
  url       = {https://proceedings.mlr.press/v70/amos17b.html}}

@article{flaschel2021unsupervised,
  author  = {Flaschel, Moritz and Kumar, Siddhant and De Lorenzis, Laura},
  title   = {Unsupervised discovery of interpretable hyperelastic constitutive laws},
  journal = {Computer Methods in Applied Mechanics and Engineering},
  volume  = {381},
  pages   = {113852},
  year    = {2021},
  doi     = {10.1016/j.cma.2021.113852}}

@article{joshi2022bayesian,
  author  = {Joshi, Akshay and Thakolkaran, Prakash and Zheng, Yiwen and Escande, Maxime and Flaschel, Moritz and De Lorenzis, Laura and Kumar, Siddhant},
  title   = {Bayesian-{EUCLID}: Discovering hyperelastic material laws with uncertainties},
  journal = {Computer Methods in Applied Mechanics and Engineering},
  volume  = {398},
  pages   = {115225},
  year    = {2022},
  doi     = {10.1016/j.cma.2022.115225}}

@article{linka2023new,
  author  = {Linka, Kevin and Kuhl, Ellen},
  title   = {A new family of {C}onstitutive {A}rtificial {N}eural {N}etworks towards automated model discovery},
  journal = {Computer Methods in Applied Mechanics and Engineering},
  volume  = {403},
  pages   = {115731},
  year    = {2023},
  doi     = {10.1016/j.cma.2022.115731}}

@article{linka2025,
  author  = {Linka, Kevin and Holzapfel, Gerhard A. and Kuhl, Ellen},
  title   = {Discovering uncertainty: {B}ayesian constitutive artificial neural networks},
  journal = {Computer Methods in Applied Mechanics and Engineering},
  volume  = {433},
  pages   = {117517},
  year    = {2025},
  doi     = {10.1016/j.cma.2024.117517}}

@article{mcculloch2025discovering,
  author  = {McCulloch, Jeremy A. and Kuhl, Ellen},
  title   = {Discovering uncertainty: {G}aussian constitutive neural networks with correlated weights},
  journal = {Computational Mechanics},
  volume  = {76},
  number  = {6},
  pages   = {1527--1539},
  year    = {2025},
  doi     = {10.1007/s00466-025-02660-y}}

@article{yi2025single,
  author  = {Yi, Jiaxiang and Ferreira, Bernardo P. and Bessa, Miguel A.},
  title   = {Single-to-multi-fidelity history-dependent learning with uncertainty quantification and disentanglement: {A}pplication to data-driven constitutive modeling},
  journal = {Computer Methods in Applied Mechanics and Engineering},
  volume  = {448},
  pages   = {118479},
  year    = {2026},
  doi     = {10.1016/j.cma.2025.118479}}

@article{padmanabha2024improving,
  author  = {Padmanabha, Govinda Anantha and Fuhg, Jan Niklas and Safta, Cosmin and Jones, Reese E. and Bouklas, Nikolaos},
  title   = {Improving the performance of {S}tein variational inference through extreme sparsification of physically-constrained neural network models},
  journal = {Computer Methods in Applied Mechanics and Engineering},
  volume  = {432},
  pages   = {117359},
  year    = {2024},
  doi     = {10.1016/j.cma.2024.117359}}

@article{casey2025exploring,
  author  = {Casey, James J. and Forsyth, Jessica E. and Waite, Timothy W. and Cotter, Simon L. and Shearer, Tom},
  title   = {Exploring natural variation in tendon constitutive parameters via {B}ayesian data selection and mixed effects models},
  journal = {Proceedings of the Royal Society A: Mathematical, Physical and Engineering Sciences},
  volume  = {481},
  number  = {2323},
  pages   = {20250034},
  year    = {2025},
  doi     = {10.1098/rspa.2025.0034}}

@article{pensalfini2023mechano,
  author  = {Pensalfini, Marco and Buganza Tepole, Adrian},
  title   = {Mechano-biological and bio-mechanical pathways in cutaneous wound healing},
  journal = {PLOS Computational Biology},
  volume  = {19},
  number  = {3},
  pages   = {e1010902},
  year    = {2023},
  doi     = {10.1371/journal.pcbi.1010902}}

@article{tac2024generative,
  author  = {Ta{\c{c}}, Vahidullah and Rausch, Manuel K. and Bilionis, Ilias and Sahli Costabal, Francisco and Buganza Tepole, Adrian},
  title   = {Generative hyperelasticity with physics-informed probabilistic diffusion fields},
  journal = {Engineering with Computers},
  volume  = {41},
  number  = {1},
  pages   = {51--69},
  year    = {2025},
  doi     = {10.1007/s00366-024-01984-2}}

@article{han2026constitutive,
  author  = {Han, Jinkyo and Bahmani, Bahador},
  title   = {Constitutive priors for inverse design},
  journal = {Computer Methods in Applied Mechanics and Engineering},
  volume  = {463},
  pages   = {119345},
  year    = {2027},
  doi     = {10.1016/j.cma.2026.119345}}

@article{bahmani2026conformal,
  author  = {Bahmani, Bahador},
  title   = {Conformal quantile regression for neural probabilistic constitutive modeling},
  journal = {Computer Methods in Applied Mechanics and Engineering},
  volume  = {457},
  pages   = {118981},
  year    = {2026},
  doi     = {10.1016/j.cma.2026.118981}}

@inproceedings{wortwein2023neural,
  author    = {W{\"o}rtwein, Torsten and Allen, Nicholas B. and Sheeber, Lisa B. and Auerbach, Randy P. and Cohn, Jeffrey F. and Morency, Louis-Philippe},
  title     = {Neural mixed effects for nonlinear personalized predictions},
  booktitle = {Proceedings of the 25th International Conference on Multimodal Interaction ({ICMI} '23)},
  publisher = {ACM},
  pages     = {445--454},
  year      = {2023},
  doi       = {10.1145/3577190.3614115}}

@article{tran2018bayesian,
  author  = {Tran, Minh-Ngoc and Nguyen, Nghia and Nott, David and Kohn, Robert},
  title   = {Bayesian deep net {GLM} and {GLMM}},
  journal = {Journal of Computational and Graphical Statistics},
  volume  = {29},
  number  = {1},
  pages   = {97--113},
  year    = {2020},
  doi     = {10.1080/10618600.2019.1637747}}

@misc{slavutsky2026neural,
  author       = {Slavutsky, Yuli and Salazar, Sebastian and Blei, David M.},
  title        = {Neural generalized mixed-effects models},
  howpublished = {arXiv preprint arXiv:2604.10976},
  year         = {2026},
  doi          = {10.48550/arXiv.2604.10976}}

@inproceedings{snoek2015scalable,
  author    = {Snoek, Jasper and Rippel, Oren and Swersky, Kevin and Kiros, Ryan and Satish, Nadathur and Sundaram, Narayanan and Patwary, Mostofa and Prabhat and Adams, Ryan},
  title     = {Scalable {B}ayesian optimization using deep neural networks},
  booktitle = {Proceedings of the 32nd International Conference on Machine Learning},
  editor    = {Bach, Francis and Blei, David},
  series    = {Proceedings of Machine Learning Research},
  volume    = {37},
  pages     = {2171--2180},
  publisher = {PMLR},
  year      = {2015}}

@incollection{harrison2018meta,
  author    = {Harrison, James and Sharma, Apoorva and Pavone, Marco},
  title     = {Meta-learning priors for efficient online {B}ayesian regression},
  booktitle = {Algorithmic Foundations of Robotics {XIII}},
  series    = {Springer Proceedings in Advanced Robotics},
  publisher = {Springer International Publishing},
  pages     = {318--337},
  year      = {2020},
  doi       = {10.1007/978-3-030-44051-0_19}}

@article{niestrawska2016microstructure,
  author  = {Niestrawska, Justyna A. and Viertler, Christian and Regitnig, Peter and Cohnert, Tina U. and Sommer, Gerhard and Holzapfel, Gerhard A.},
  title   = {Microstructure and mechanics of healthy and aneurysmatic abdominal aortas: experimental analysis and modelling},
  journal = {Journal of the Royal Society Interface},
  volume  = {13},
  number  = {124},
  pages   = {20160620},
  year    = {2016},
  doi     = {10.1098/rsif.2016.0620}}

@book{holzapfel2000nonlinear,
  author    = {Holzapfel, Gerhard A.},
  title     = {Nonlinear Solid Mechanics: A Continuum Approach for Engineering},
  publisher = {John Wiley \& Sons},
  isbn      = {978-0-471-82319-3},
  year      = {2000}}

@article{treloar1944stress,
  author  = {Treloar, Leslie R. G.},
  title   = {Stress-strain data for vulcanised rubber under various types of deformation},
  journal = {Transactions of the Faraday Society},
  volume  = {40},
  pages   = {59--70},
  year    = {1944},
  doi     = {10.1039/TF9444000059}}

@article{simpson2017penalising,
  author  = {Simpson, Daniel and Rue, H{\aa}vard and Riebler, Andrea and Martins, Thiago G. and S{\o}rbye, Sigrunn H.},
  title   = {Penalising model component complexity: A principled, practical approach to constructing priors},
  journal = {Statistical Science},
  volume  = {32},
  number  = {1},
  pages   = {1--28},
  year    = {2017},
  doi     = {10.1214/16-STS576}}

@article{egozcue2003isometric,
  author  = {Egozcue, Juan Jos{\'e} and Pawlowsky-Glahn, Vera and Mateu-Figueras, Gloria and Barcel{\'o}-Vidal, Carles},
  title   = {Isometric logratio transformations for compositional data analysis},
  journal = {Mathematical Geology},
  volume  = {35},
  number  = {3},
  pages   = {279--300},
  year    = {2003},
  doi     = {10.1023/A:1023818214614}}

@book{gelman2006data,
  author    = {Gelman, Andrew and Hill, Jennifer},
  title     = {Data Analysis Using Regression and Multilevel/Hierarchical Models},
  publisher = {Cambridge University Press},
  address   = {Cambridge},
  year      = {2006},
  doi       = {10.1017/CBO9780511790942}}

@article{gelman2006prior,
  author  = {Gelman, Andrew},
  title   = {Prior distributions for variance parameters in hierarchical models},
  journal = {Bayesian Analysis},
  volume  = {1},
  number  = {3},
  pages   = {515--534},
  year    = {2006},
  doi     = {10.1214/06-BA117A}}

@incollection{lecun1998efficient,
  author    = {LeCun, Yann and Bottou, L{\'e}on and Orr, Genevieve B. and M{\"u}ller, Klaus-Robert},
  title     = {Efficient {BackProp}},
  booktitle = {Neural Networks: Tricks of the Trade},
  editor    = {Orr, Genevieve B. and M{\"u}ller, Klaus-Robert},
  series    = {Lecture Notes in Computer Science},
  volume    = {1524},
  pages     = {9--50},
  publisher = {Springer},
  address   = {Berlin, Heidelberg},
  year      = {1998},
  doi       = {10.1007/3-540-49430-8_2}}

@inproceedings{hoedt2023principled,
  author    = {Hoedt, Pieter-Jan and Klambauer, G{\"u}nter},
  title     = {Principled weight initialisation for input-convex neural networks},
  booktitle = {Advances in Neural Information Processing Systems},
  volume    = {36},
  pages     = {46093--46104},
  year      = {2023},
  doi       = {10.52202/075280-1998}}

@article{papaspiliopoulos2007general,
  author  = {Papaspiliopoulos, Omiros and Roberts, Gareth O. and Sk{\"o}ld, Martin},
  title   = {A general framework for the parametrization of hierarchical models},
  journal = {Statistical Science},
  volume  = {22},
  number  = {1},
  pages   = {59--73},
  year    = {2007},
  doi     = {10.1214/088342307000000014}}

@article{ball1977convexity,
  author  = {Ball, John M.},
  title   = {Convexity conditions and existence theorems in nonlinear elasticity},
  journal = {Archive for Rational Mechanics and Analysis},
  volume  = {63},
  number  = {4},
  pages   = {337--403},
  year    = {1976},
  doi     = {10.1007/BF00279992}}

@inproceedings{kingma2014adam,
  author    = {Kingma, Diederik P. and Ba, Jimmy},
  title     = {Adam: A method for stochastic optimization},
  booktitle = {3rd International Conference on Learning Representations ({ICLR} 2015), San Diego, CA, USA, Conference Track Proceedings},
  year      = {2015},
  note      = {arXiv:1412.6980},
  doi       = {10.48550/arXiv.1412.6980}}

@article{hoffman2014nuts,
  author  = {Hoffman, Matthew D. and Gelman, Andrew},
  title   = {The {No-U-Turn} sampler: adaptively setting path lengths in {Hamiltonian Monte Carlo}},
  journal = {Journal of Machine Learning Research},
  volume  = {15},
  number  = {47},
  pages   = {1593--1623},
  year    = {2014}}

@article{yin2023generative,
  author  = {Yin, Minglang and Zou, Zongren and Zhang, Enrui and Cavinato, Cristina and Humphrey, Jay D. and Karniadakis, George Em},
  title   = {A generative modeling framework for inferring families of biomechanical constitutive laws in data-sparse regimes},
  journal = {Journal of the Mechanics and Physics of Solids},
  volume  = {181},
  pages   = {105424},
  year    = {2023},
  doi     = {10.1016/j.jmps.2023.105424}}

@book{bishop2024deep,
  author    = {Bishop, Christopher M. and Bishop, Hugh},
  title     = {Deep Learning: Foundations and Concepts},
  publisher = {Springer},
  address   = {Cham},
  year      = {2024},
  doi       = {10.1007/978-3-031-45468-4}}

@article{lindstrom1990nonlinear,
  author  = {Lindstrom, Mary J. and Bates, Douglas M.},
  title   = {Nonlinear mixed effects models for repeated measures data},
  journal = {Biometrics},
  volume  = {46},
  number  = {3},
  pages   = {673--687},
  year    = {1990},
  doi     = {10.2307/2532087}}

@article{klein2023parametrised,
  author  = {Klein, Dominik K. and Roth, Fabian J. and Valizadeh, Iman and Weeger, Oliver},
  title   = {Parametrized polyconvex hyperelasticity with physics-augmented neural networks},
  journal = {Data-Centric Engineering},
  volume  = {4},
  pages   = {e25},
  year    = {2023},
  doi     = {10.1017/dce.2023.21}}

@article{hoffman2013stochastic,
  author  = {Hoffman, Matthew D. and Blei, David M. and Wang, Chong and Paisley, John},
  title   = {Stochastic variational inference},
  journal = {Journal of Machine Learning Research},
  volume  = {14},
  number  = {40},
  pages   = {1303--1347},
  year    = {2013}
}

@article{blei2017variational,
  author  = {Blei, David M. and Kucukelbir, Alp and McAuliffe, Jon D.},
  title   = {Variational inference: A review for statisticians},
  journal = {Journal of the American Statistical Association},
  volume  = {112},
  number  = {518},
  pages   = {859--877},
  year    = {2017},
  doi     = {10.1080/01621459.2017.1285773}
}

@inproceedings{kingma2014autoencoding,
  author    = {Kingma, Diederik P. and Welling, Max},
  title     = {Auto-encoding variational {B}ayes},
  booktitle = {2nd International Conference on Learning Representations ({ICLR} 2014), Banff, AB, Canada, Conference Track Proceedings},
  year      = {2014},
  note      = {arXiv:1312.6114},
  doi       = {10.48550/arXiv.1312.6114}
}

@misc{holthusen2026benchmarking,
  author       = {Holthusen, Hagen and Flaschel, Moritz and Martonov{\'a}, Denisa and Kuhl, Ellen},
  title        = {Benchmarking data-driven material models on the classic {T}reloar dataset},
  howpublished = {arXiv preprint arXiv:2608.14063},
  year         = {2026},
  doi          = {10.48550/arXiv.2608.14063}}

@article{ghaboussi1991knowledge,
  author  = {Ghaboussi, J. and Garrett, J. H. and Wu, X.},
  title   = {Knowledge-based modeling of material behavior with neural networks},
  journal = {Journal of Engineering Mechanics},
  volume  = {117},
  number  = {1},
  pages   = {132--153},
  year    = {1991},
  doi     = {10.1061/(ASCE)0733-9399(1991)117:1(132)}}

@article{fuhg2022local,
  author  = {Fuhg, Jan N. and Marino, Michele and Bouklas, Nikolaos},
  title   = {Local approximate {G}aussian process regression for data-driven constitutive models: development and comparison with neural networks},
  journal = {Computer Methods in Applied Mechanics and Engineering},
  volume  = {388},
  pages   = {114217},
  year    = {2022},
  doi     = {10.1016/j.cma.2021.114217}}

@article{wang2021inference,
  author  = {Wang, Zhenlin and Estrada, Jonathan B. and Arruda, Ellen M. and Garikipati, Krishna},
  title   = {Inference of deformation mechanisms and constitutive response of soft material surrogates of biological tissue by full-field characterization and data-driven variational system identification},
  journal = {Journal of the Mechanics and Physics of Solids},
  volume  = {153},
  pages   = {104474},
  year    = {2021},
  doi     = {10.1016/j.jmps.2021.104474}}

@article{flaschel2026material,
  author  = {Flaschel, Moritz and Martonov{\'a}, Denisa and Veil, Carina and Kuhl, Ellen},
  title   = {Material {F}ingerprinting: a shortcut to material model discovery without solving optimization problems},
  journal = {Computer Methods in Applied Mechanics and Engineering},
  volume  = {450},
  pages   = {118573},
  year    = {2026},
  doi     = {10.1016/j.cma.2025.118573}}

@article{flaschel2026unsupervised,
  author  = {Flaschel, Moritz and Moreno-Mateos, Miguel Angel and Wiesheier, Simon and Steinmann, Paul and Kuhl, Ellen},
  title   = {Unsupervised material fingerprinting: ultra-fast hyperelastic model discovery from full-field experimental measurements},
  journal = {Computer Methods in Applied Mechanics and Engineering},
  volume  = {461},
  pages   = {119256},
  year    = {2026},
  doi     = {10.1016/j.cma.2026.119256}}

@article{flaschel2026adaptive,
  author  = {Flaschel, Moritz and Holthusen, Hagen and Martonov{\'a}, Denisa and Kuhl, Ellen},
  title   = {Adaptive {M}aterial {F}ingerprinting for the fast discovery of polyconvex feature combinations in isotropic and anisotropic hyperelasticity},
  journal = {International Journal of Engineering Science},
  volume  = {227},
  pages   = {104594},
  year    = {2026},
  doi     = {10.1016/j.ijengsci.2026.104594}}

@article{abdusalamov2023automatic,
  author  = {Abdusalamov, Rasul and Hillg{\"a}rtner, Markus and Itskov, Mikhail},
  title   = {Automatic generation of interpretable hyperelastic material models by symbolic regression},
  journal = {International Journal for Numerical Methods in Engineering},
  volume  = {124},
  number  = {9},
  pages   = {2093--2104},
  year    = {2023},
  doi     = {10.1002/nme.7203}}

@article{tacke2025constitutive,
  author  = {Tacke, Marius and Busch, Matthias and Bali, Kartik and Abdolazizi, Kian and Linka, Kevin and Cyron, Christian and Aydin, Roland},
  title   = {Constitutive scientific generative agent ({CSGA}): leveraging large language models for automated constitutive model discovery},
  journal = {Machine Learning for Computational Science and Engineering},
  volume  = {1},
  number  = {1},
  pages   = {23},
  year    = {2025},
  doi     = {10.1007/s44379-025-00022-2}}

@article{wiesheier2024versatile,
  author  = {Wiesheier, Simon and Moreno-Mateos, Miguel Angel and Steinmann, Paul},
  title   = {Versatile data-adaptive hyperelastic energy functions for soft materials},
  journal = {Computer Methods in Applied Mechanics and Engineering},
  volume  = {430},
  pages   = {117208},
  year    = {2024},
  doi     = {10.1016/j.cma.2024.117208}}

@article{scholz2026thermoplastic,
  author  = {Scholz, Anneke and Selle, Ricky and Gro{\ss}hauser, Michael},
  title   = {Thermoplastic post-consumer recyclates for buried infrastructure: barriers, classification gaps, and a path toward quality-assured use},
  journal = {Polymers},
  volume  = {18},
  number  = {14},
  pages   = {1702},
  year    = {2026},
  doi     = {10.3390/polym18141702}}

@article{tipping2001sparse,
  author  = {Tipping, Michael E.},
  title   = {Sparse {B}ayesian learning and the relevance vector machine},
  journal = {Journal of Machine Learning Research},
  volume  = {1},
  pages   = {211--244},
  year    = {2001}}

@article{anton2026uncertainty,
  title   = {Uncertainty quantification in model discovery by distilling interpretable material constitutive models from {G}aussian process posteriors},
  author  = {Anton, David and Wessels, Henning and R{\"o}mer, Ulrich and Henkes, Alexander and Urrea-Quintero, Jorge-Humberto},
  journal = {Computer Methods in Applied Mechanics and Engineering},
  volume  = {458},
  pages   = {118979},
  year    = {2026},
  doi     = {10.1016/j.cma.2026.118979},
  publisher = {Elsevier}
}

@article{szekely2013energy,
  author  = {Sz{\'e}kely, G{\'a}bor J. and Rizzo, Maria L.},
  title   = {Energy statistics: A class of statistics based on distances},
  journal = {Journal of Statistical Planning and Inference},
  volume  = {143},
  number  = {8},
  pages   = {1249--1272},
  year    = {2013},
  doi     = {10.1016/j.jspi.2013.03.018}}

@article{lewandowski2009generating,
  author  = {Lewandowski, Daniel and Kurowicka, Dorota and Joe, Harry},
  title   = {Generating random correlation matrices based on vines and extended onion method},
  journal = {Journal of Multivariate Analysis},
  volume  = {100},
  number  = {9},
  pages   = {1989--2001},
  year    = {2009},
  doi     = {10.1016/j.jmva.2009.04.008}}

@incollection{platt1999probabilistic,
  author    = {Platt, John C.},
  title     = {Probabilistic outputs for support vector machines and comparisons to regularized likelihood methods},
  booktitle = {Advances in Large Margin Classifiers},
  editor    = {Smola, Alexander J. and Bartlett, Peter L. and Sch{\"o}lkopf, Bernhard and Schuurmans, Dale},
  publisher = {MIT Press},
  address   = {Cambridge, MA},
  pages     = {61--74},
  year      = {2000}
}

\end{document}